\documentclass[a4paper, 10pt, twocolumn]{article}
\usepackage[utf8x]{inputenc}
\usepackage{amsthm}
\usepackage{amsmath}
\usepackage{amssymb}
\usepackage{subfigure}
\usepackage{epsfig}
\usepackage{algorithmic}
\usepackage{algorithm} 
\usepackage{color}
\usepackage{authblk}
\usepackage{hyperref}

\newcommand*\oldc{}\let\oldc\c

\renewcommand{\c}{\textbf{c}}
\renewcommand{\d}{\textbf{d}}

\newcommand{\x}{\textbf{x}}

\newcommand{\A}{\textbf{A}}

\renewcommand*\c{}\let\c\oldc 

\newtheorem{definition}{Definition}[section]

\begin{document}

\title{The Evolution of the Cuban HIV/AIDS Network}
\author[1, *]{Charanpal Dhanjal }
\author[1]{St\'{e}phan Cl\'{e}men\c{c}on}
\author[2, 3]{Hector De Arazoza} 
\author[1]{Fabrice Rossi} 
\author[3, 4]{Viet Chi Tran} 

\affil[1]{Institut Telecom, LTCI UMR Telecom ParisTech/CNRS No. 5141, 46 rue Barrault, 75634 Paris Cedex 13, France}
\affil[2]{Facultad de Matem\'atica y Computaci\'on, Universidad de la Habana, La Habana, Cuba}
\affil[3]{Laboratoire Paul Painlev\'e, UMR CNRS 8524, Universit\'e Lille 1, 59 655 Villeneuve d'Ascq Cedex, France}
\affil[4]{ Centre de Math\'{e}matiques Appliqu\'{e}es, UMR CNRS 7641, Ecole Polytechnique, Route de Saclay, 91 128 Palaiseau Cedex, France.}

\date{\today}
\maketitle

\pagestyle{plain}

\maketitle

\let\oldthefootnote\thefootnote
\renewcommand{\thefootnote}{\fnsymbol{footnote}}
\footnotetext[1]{Author for correspondence (dhanjal@telecom-paristech.fr)}
\let\thefootnote\oldthefootnote

%



\begin{abstract} 
An individual detected as HIV positive in Cuba is asked to provide a list of his/her sexual contacts for the previous 2 years. This allows one to gather detailed information on the spread of the HIV epidemic. Here we study the evolution of the sexual contact graph of detected individuals and also the directed graph of HIV infections. The study covers the Cuban HIV epidemic between the years 1986 and 2004 inclusive and is motivated by an earlier study on the static properties of the network at the end of 2004. We use a variety of advanced graph algorithms to paint a picture of the growth of the epidemic, including an examination of diameters, geodesic distances, community structure and centrality amongst others characteristics. The analysis contrasts the HIV network with other real networks, and graphs generated using the configuration model. We find that the early epidemic starts in the heterosexual population and then grows mainly through MSM (Men having Sex with Men) contact. The epidemic exhibits a giant component which is shown to have degenerate chains of vertices and after 1989, diameters are larger than that expected by the equivalent configuration model graphs. In 1997 there is an significant increase in the detection rate from 73 to 256 detections/year covering mainly MSMs which results in a rapid increase of distances and diameters in the giant component.
\end{abstract}



\section{Introduction}

In Cuba, the HIV epidemic is recorded and controlled using detections made in a number of different ways, including \emph{contact tracing} which is the process of finding and testing the contacts of a person detected as HIV positive. This detection system, set up in 1986, has been effective at keeping the HIV epidemic under control and has also enabled the recording of a considerable amount of epidemiological data. Any individual tested as HIV positive is indexed and described through several attributes including gender, age, sexual preference, way of detection and area of residence. Furthermore, a list of indices corresponding to the sexual partners in the database she/he possibly named is available, allowing one to construct a graph of sexual contacts along which the epidemic spreads. Note that HIV most often spreads in Cuba through sexual transmission, and infection cases by blood transfusion or injection drug use have negligible impact on the evolution of the epidemic. In many situations, a medical investigation allows one to obtain a plausible date of infection together with the index of the supposed HIV transmitter.

A preliminary overview of the HIV/AIDS epidemics in Cuba is provided in \cite{Auvert07} and the characteristics of the network at the end of 2004 are studied in \cite{clemencondearazozarossitran}. One of the conclusions of the study of \cite{clemencondearazozarossitran} is that it is necessary to analyse the time varying characteristics of the HIV epidemic. Hence we use recent graph-mining techniques and statistical analysis to learn how the network has evolved between the start of 1986 and the end of 2004. We aim to describe the connectivity and communication properties of the network and understand the impact of heterogeneity (with respect to the attributes observed) in the graph structure and its evolution. We also compare the epidemic graph to patterns observed in a variety of other real networks and a set of generated networks with matching degrees to the real epidemic.  

Our database contains 1109 women and 4280 men, a total of 5389 individuals and 4097 edges. Of this set 2122 people died by the year 2006. As a preprocessing step on the graph of sexual contacts, we take the union of infection and contact edges. This ensures that if an infection is recorded from one individual from another then an edge exists between the corresponding vertices for the sexual contact graph. Notice that due to the nature of the way data is recorded, gender and sexual orientation can be merged into the groups women, heterosexual men, and MSM (Men having Sex with Men: those who reported at least one sexual contact with another man in the two years preceding HIV detection). Note also that contact tracing starts not immediately after a person gives her/his sexual partners (for the last two years), but takes some time. When a person is found to be HIV positive, there is an epidemiological interview, carried out by the Epidemiology Department of the municipality where the person is residing or by the Regional Sanatorium for HIV/AIDS. Then the contacts the person gives are forwarded to the corresponding municipalities where they reside. This takes time, therefore some people that should be in the contact tracing ``road'' to detection are detected some time after the primary case (the one that gives him/her as contact) has been detected.

To discuss the properties we will later examine, we first formalise the representation of the epidemic graph (for an introduction to the study of graphs, see \cite{newman03graphs, durrett, chunglu,vanderhofstad}). The \emph{contact graph} at time $t$ is $G_t = (V_t, E_t)$. Here, $V_t = \{v_1, \ldots, v_{|V_t|}\}$ is a set of vertices, and $E_t \subseteq V_t \times V_t$ is a set of edges where $|V_t|$ is the number of vertices at time $t$. Each vertex represents an individual detected as having HIV and edges between vertices represent recorded sexual contact. Since edges and vertices are not deleted with time, $V_t \subseteq V_{t+1}$ and $E_t \subseteq E_{t+1}$ for all times $t$. Furthermore, our graph is undirected which means that if $(v_i, v_j)$ is an edge then $(v_j, v_i)$ is also an edge. As we also have information on the possible infections we can construct an \emph{infection graph} which has directed edges $D_t \subseteq V_t \times V_t$ to indicate infections. The set of edges representing infections at time $t$ is a subset of the set of sexual contact edges, i.e. $D_t \subseteq E_t$, since we assume an infection can only occur between individuals who have had sexual contact.

We begin with some background to contextualise this work in terms of the analysis of real and random graph growth, and the study of epidemics. Following, in Section \ref{sec:aimsMethods} we outline the main aims beyond a temporal exploration of the data, and also the methods used for analysis. Section \ref{sec:resultsAnalysis} presents the resulting analysis on the epidemic data and in the final part of this paper we summarise findings and provide concluding remarks. 

\section{Background} 

The \emph{degree} of a vertex in an undirected graph is the number of edges incident to it, and in a directed graph one can talk about the number of outgoing and incoming edges of a vertex as the out-degrees and in-degrees respectively. An early observation on the characteristics of real graphs is that the \emph{degree distribution} follows a particular pattern.  The degree distribution gives the probabilities $(p(k))_{k\geq 0}$ that a vertex selected at random has a degree $k \geq 0$. In many real-world networks (e.g. \cite{jeong2001lethality, redner1998popular, chen2002origin, albert1999internet})  the degree distribution follows a \emph{power law}, which means that $p(k) \propto k^{-\alpha}$ for some exponent $\alpha > 1$. This means that many vertices have few connections but a small fraction have large degrees. One of the first network growth models to show a power law distribution was the Albert-Barabasi model \cite{barabasi1999emergence}. In this model vertices are added incrementally and when a new vertex is added to the network, it is linked to an existing vertex with a probability proportional to the degree. The degree exponent in the Albert-Barabasi model is proven to be $3$. Modifications of the preferential attachment model have been studied by many authors, for instance in the ageing of nodes where the probability to connect to a node decreases with its age and this is shown to change the exponent of the power distribution. See also for example the works by Dorogovtsev and Mendes \cite{dorogovtsevmendes,dorogovtsevmendes2000,dorogovtsevmendes2001} and by Kumar and co-authors \cite{kleinbergetal,kumaretal2000}.

There have been a number of other random graph models which possess a power law degree distribution. In \cite{saramaki2004scale} a random graph is constructed with a degree exponent of 3 using random walks.  Another random graph model which generates a power law degree distribution is the copying model \cite{kleinberg99web} which also results in a community structure. For a detailed review of these models and their variations, we refer the reader to \cite{mitzenmacher2004brief}. One way of generating a graph with given degree distribution, including the power law distribution, is provided by the configuration model (CM, \cite{bollobas2001,durrett,molloyreed, vanderhofstad}). In this model, for each vertex $v_i$ a degree is sampled from a given degree distribution and one can consider the vertices as having the corresponding number of ``spokes'' sticking out of them. One then randomly pairs unconnected spokes in order to form a graph under the required degree distribution.

\subsection{Real Graph Growth}

Of particular relevance to the present study is the study of the sexual contact network in Sweden \cite{liljeros2001web} in which 4781 individuals were surveyed about their sexual behaviour. Using the graph constructed from the sexual contacts up to 12 months before the survey, the authors noted a power law exponent of 2.4 for males and females. Two reasons accounting for the power law behaviour are: increased skill in getting new partners as the number of previous partners grows and different levels of attractiveness. In \cite{jones2003assessment} a similar study is conducted using heterosexual contact networks in Sweden, Uganda and the USA and by fitting to the Yule distribution.

The growth characteristics of real networks have been studied in a wide variety of domains. A review on the evolution of networks is presented in \cite{dorogovtsev2001evolution, dorogovtsevmendes}. In these works the authors detail a number of studies on citation and collaboration networks, networks of internet routers, neural and protein networks and electronic circuits amongst others.  Further interesting studies include \cite{barabasi02socialEv} on the evolution of a scientific collaboration network, \cite{kossinets06socialEv} which studies a university social network, \cite{holmes04structure} which considers an Internet dating community and \cite{mislove2008growth} which examines 3 months of growth in the Flickr social network.

In Leskovec et al. \cite{leskovec2005graphs, leskovec2007graph} the evolution of a number of real graphs are studied over time. The graphs considered include citation graphs for articles and patents, a graph of routers within the Internet, and a set of bipartite affiliation graphs of papers and authors. The authors find that the networks become denser with time (vertices increase their out-degree with time), encapsulated in the so-called Densification Power Law (DPL). Furthermore the diameter, defined as the longest shortest path, decreases indicating a small-world effect in which a large network has a low average path length between pairs of vertices

The DPL was independently observed in a citation network in Katz \cite{katz2005scale}. It states that the number of edges grows exponentially faster than the number of vertices, $|E_t| \propto |V_t|^\sigma$ where $|E_t|$ and $|V_t|$ are respectively the number of edges and vertices at time $t$, and the value of $\sigma$ is often between $1$ and $2$. In order to better model these effects \cite{leskovec2005graphs} proposes two graph models: the community guided attachment and the forest fire graph. This later model generates both the densification property and a shrinking diameter. Another graph growth model which satisfies these properties is the Kronecker graph \cite{leskovec2005realistic, leskovec2010kronecker}, which is based on successive Kronecker multiplication of adjacency matrices.

\subsection{Epidemic Networks} 

There is a considerable amount of research which studies stochastic models of epidemic networks and their associated properties, and here we briefly recount some interesting results. The interested reader is directed to \cite{andersson2000stochastic} for a detailed review. A well-known compartmentalized epidemic model is the Susceptible Infected Removed (SIR) one which was introduced in \cite{kermack27epidemics}. A set of differential equations describes how these groups of individuals change over time.  The spread of infections on spatial lattices have been for instance considered by \cite{kleczkowski99meanfield}. Similar studies involving mean-field approximations are considered in \cite{barthelemybarratpastorsatorrasvespignani,durrett,pastorsatorrasvespignani}.

The small world property of real networks is important in the spread of disease as it implies that paths between pairs of vertices can be small and the spread of an infectious disease could potentially spread quickly over a population. Studies involving disease transmission over a small world network are treated by Ball Neal \cite{ballneal2003} and Moore and Newman \cite{moore00epidemics}. 

Propagation of epidemics on CM graphs has been considered for instance by Ball and Neal \cite{ballneal}. Further work along these lines include those of Volz and Meyers \cite{volz,volzmeyers}, Decreusefond et al. \cite{decreusefonddhersinmoyaltran} where approximations by ordinary differential equations have been proposed in the case of large epidemics. In Ball, Sirl and Trapman \cite{ballsirltrapman2009,ballsirltrapman2010} the authors propose a configuration graph which also has the small world property of high clustering motivated in part due to a ``household'' structure. 

In line with this work is the notion of a random intersection graph, introduced by \cite{karonskischeinermansingercohen}. These graphs have been used for epidemiology by Britton et al. \cite{brittondeijfenlageraslindholm}. The authors investigate the effect of the clustering within the graph on the epidemic threshold. 

\section{Aims and Method} \label{sec:aimsMethods}

As previously mentioned, a study has already been conducted on the static graph characteristics of the epidemic and here we extend this analysis in a temporal sense in order to understand how the epidemic evolves. The main aim is to perform an exploration of the properties of the data, however, in addition we would like to answer the following questions: 

\begin{enumerate} 
 \item How do the properties of a set of growing CM graphs generated with degree distributions of the real epidemic vary from those seen in the evolving HIV graph?
 \item How do the different detection methods change their roles over time? 
 \item Does the evolution of the epidemic network follow the same properties observed in \cite{leskovec2005graphs, leskovec2007graph}? 
\end{enumerate}

The first question identifies differences in the characteristics of the real epidemic graph with a set of random graphs with the same degree distribution over time. In order to conduct the analysis we generate 10 CM random graphs with specified degree distributions identical to that of the HIV epidemic at 3 month intervals. Resulting characteristics are averaged over all random graphs in order to make a comparison with the equivalent characteristics on the real graph. This analysis is conducted on both the sexual contact graph and infection graph. In the case of the infection graph, the CM is modified to generate directed graphs and one changes the algorithm in an intuitive fashion to allow both the in-degrees and out-degrees.

For the exploratory analysis and in order to address the remaining questions, we compute statistics over both contact and infection graphs which are described in the follow subsections. In most cases graph properties are evaluated 90 day intervals starting from 03/01/86 and ending on 02/10/04. For some properties the quantities of interest are recorded at 4 year intervals in the same date range, namely the dates  31/12/89, 31/12/93, 31/12/97, 31/12/01 and 02/10/04. Most of the analysis code is written in Python using the APGL graph library (http://packages.python.org/apgl) however clustering is performed using the igraph package for R (http://igraph.sourceforge.net). 

\subsection{Global Properties and Connectivity}

We begin with analyses which gives a global picture of the evolving structure of the epidemic, and plot the number of vertices and edges in each graph. To test if the DPL is exemplified in our data, we plot  $\log(|E_t|)$ and $\log(|D_t|)$ versus $\log(|V_t|)$ for which the slope is the exponent $\sigma$. In the datasets used in \cite{leskovec2007graph}, the values of the exponent fall between $1$ and $2$. An exponent of 1 corresponds to a constant average degree over time and $\sigma = 2$ implies that on average each vertex has edges to a constant fraction of all vertices.  

Following, we perform a study of the components in the graph. For a graph $G$ two vertices are said to be \emph{connected} if there is a path between them. In the same way a graph is said to be connected if all pairs of vertices are connected. The HIV epidemic graph is not connected. However, one can talk about \emph{connected components} of the graph i.e. sets of vertices which are connected. The component distribution is recorded at 4 year intervals for both the contact and infection graph (in the latter we consider the distribution of trees), as is the number of components of different sizes. As connectivity is related to the detection methods we explore their evolution over time and also the total number of sexual contacts declared by each individual, as well as the number tested and found positive. 

We also study the mean degrees of the graphs to make the comparison to \cite{leskovec2005graphs, leskovec2007graph} in which mean degrees are shown to always increase. Following, we examine the degree change at an individual level by considering the difference in the days between two individuals connected by an edge in the contact graph. We call the absolute value of the difference in days between detection dates of two connected persons their \emph{detection distance}. 

\subsection{The Largest Component} 

In many networks there exists a \emph{giant component} which is a component of size comparable to the entire graph. We observed a giant component at the end of the period of the recorded epidemic, and we would like to know when a giant component comes into formation and how it grows. To this effect, we record the maximum component size and its number of edges at 90 day intervals. In the infection graph we look at size evolution of the largest trees. 

A common property observed in many real networks is the \emph{small world} effect. This property states that even in large networks the shortest path between any two vertices is generally small. One way of measuring if the epidemic network has a small world property is to record the diameter of the graph. Diameter however can be subject to degenerate structures such as long chains in the graph. A more robust measure is the effective diameter, defined in the following way: 

\begin{definition}[Effective Diameter]
For each natural number $d$ and a graph $G$, let $g(d)$ denote the fraction of connected vertices which have a shortest path between them of length at most $d$. Let $D$ be an integer for which $g(D-1) < q$ and $g(D) \geq q$, for some user-defined value $q$. Then the graph $G$ has the effective diameter $D$. 
\end{definition}

In the experiments presented in \cite{leskovec2007graph}, the diameter and effective diameter exhibit similar behaviour. In particular it should be noted that effective diameter is equivalent to diameter when $q=1$. We compute effective diameters using $q=0.9$ in our experiments. Analogously, for the infection trees we study the depths of the largest trees. 

To extend the notion of measuring shortest paths within a graph, one can consider the \emph{average geodesic distance}: 

\begin{definition}[Average Geodesic Distance]\label{def:averagegeodesicdistance}
For an undirected graph $G$, let $d(v_i, v_j)$ be the distance of the shortest path between vertices $v_i$ and $v_j$, with $d(v_i, v_i) = 0$, and $d(v_i, v_j) = 0$ if $v_i$ and $v_j$ are not connected. Then the average geodesic distance is 
\begin{displaymath} 
\ell(G) = \frac{2}{|V|(|V|+1)} \sum_{i \leq j} d(v_i, v_j). 
\end{displaymath}
\end{definition}

In other words the average geodesic distance is a measure of the average shortest path length between all pairs of connected vertices.
The diameter and geodesic distance give information on the set of all shortest paths between vertices. We additionally consider a measure of the mean growth of the neighbourhood of vertices with path length using a \emph{hop plot} \cite{palmer2002anf}.
\begin{definition}[Hop plot]Let $f(h)$ be the number of pairs of vertices that are reachable within path lengths (hops) of $h$ or less, then the hop plot is a plot of $f(h)$ with $h$.
\end{definition}

\subsubsection{Community Structure} 

A way of finding the community structure of vertices is to study those subgraphs in which there are many edges within each subgraph and few between different subgraphs. As we are dealing with a loosely connected graph, it is natural to consider the giant component for this study, and this component has already been considered in \cite{clemencondearazozarossitran,clemenconarazozarossitranESANN,clemenconarazozarossitranIWANN}. We use the maximization of the modularity defined in \cite{clemencondearazozarossitran}, together with an algorithm by Noack and Rotta \cite{NoackRotta2009MultiLevelModularity} to cluster the giant component obtained in the whole data set. Notice that the graph clustering does not take into account the temporal aspect of the interactions. More precisely, we study the final graph that contains all detected individuals, regardless of their detection date. In addition, each edge of the graph (sexual contact) is considered to exist ``forever'', in a static final graph.

We measure the change in the sexual orientations of the clusters in the giant component as the epidemic evolves, comparing to the distribution of orientations in the whole population. To discover more about whether the clustering reveals interesting temporal homogeneity we also plot the temporal variation in the detection dates for several clusters. Finally, we look at the distribution of detection distances for those edges within the clusters and between clusters. 

\subsection{Gender, Sexual Orientation and Location} 

In the static analysis of the epidemic network it was clear that the gender and orientation of individuals are key factors for understanding changes in the network. Hence we study these properties as well location. We first look at genders and orientations over the whole graph, and then study the breakdown for those people detected using contact tracing and doctor recommendations. The infections along edges are also examined with respect to gender and orientation. Closely related to orientations is the study of the distribution of \textit{triangle participations} of vertices in the contact graph. For each vertex we count how many triangles that vertex participates within, which implies a set of 3 MSM vertices with edges between them. We would in general expect the triangle participation distribution to be strongly peaked at zero and have a heavy tail, however the distribution is interesting since it provides an insight into the clustering of the MSM population, which is a key facet of the epidemic spread.

Given a set of infection trees an interesting question is whether the individuals within the tree are homogeneous in terms of their location and sexual orientation. Hence we measure a quantity called the \emph{information entropy} \cite{shannon2001mathematical} of these characteristics for each tree which has a size greater than 1. Intuitively, entropy is a measure of the unpredictability of a random variable $X$. As an example a fair coin (each side lands with probability $0.5$) has an entropy of $1$, whereas an unfair one which always lands on heads has an entropy of $0$.

\begin{definition}[Entropy]Let $X$ have possible values $x_1, \ldots, x_k$, then entropy is defined as:
\begin{displaymath} 
 H_b(X) = \sum_{i=1}^k - p(x_i) \log_b(p(x_i)),
\end{displaymath}
where $b$ is the base of the logarithm and $b=2$ in our experiments.
\end{definition}

\subsection{Subgraph/subset Distances and Degrees}\label{subsection:degreeprties}

One of the problems with the average geodesic distance of Definition \ref{def:averagegeodesicdistance} is that in graphs that are not connected, some of the distances between vertices are infinite. Another way of measuring the mean distances in disconnected graphs is to use the \emph{harmonic mean geodesic distance} in which infinite distances contribute nothing to the sum:

\begin{definition}[Harmonic Average Geodesic Distance]
Let $d(v_i, v_j)$ be the distance of the shortest path between vertices $v_i$ and $v_j$. Then the harmonic average geodesic distance for an undirected graph $G$ is
\begin{displaymath}
\ell'(G)^{-1} = \frac{2}{|V|(|V|+1)} \sum_{i \leq j} d(v_i, v_j)^{-1}.
\end{displaymath}
\end{definition}

This is informative since it includes components other than the largest one. In addition we look at the distance evolution in the subgraph of just men, knowing that MSM contact is a main driver for the evolution.  This is followed by a study of the distances between the subset of MSM individuals and those with the top 10\% highest degrees.

We plot the degree distribution at various points in time for both the contact and infection graph. For the contact graph we also compute the power law exponent $\alpha$ as the graph grows. Estimation of $\alpha$ can be achieved by minimizing K\"{u}llback-Leibler divergence or by computing Hill estimators, see \cite{clemencondearazozarossitran} for more details. In our case we use K\"{u}llback-Leibler minimization. Notice that a large value of $\alpha$ implies a short tailed distribution in which large degrees are less probable. In \cite{leskovec2007graph}, for an email network the degree exponent for the \emph{power law degree distribution} is almost constant over time whereas with a citation graph the exponent decreases over time. These effects are well predicted by Theorems 5.1 and 5.2 of the paper, and we would like to know whether these theorems can predict the degree exponent in the epidemic graph.

\subsection{Centrality}

The final set of analyses concerns the \emph{centrality} of vertices which is a measure of their relative importance, and also the resilience of the graph. By resilience we mean that if we remove a central vertex then a resilient graph still has connectivity. There are several measures of centrality in existence \cite{bollobas2001} and we consider the \emph{eigenvector centrality}. Relative scores to all vertices in the network are based on the principal that connections to high-scoring vertices contribute more to the score of the vertex in question than equal connections to low-scoring nodes. This method is used as a basis of Google's PageRank \cite{brin1998anatomy} for example. Eigenvector centrality is defined more formally as: 

\begin{definition}[Eigenvector centrality]
For a graph $G$, let $\x_i$ denote the score of vertex $v_i$, and let $\A$ be the adjacency matrix of $G$ such that $\A_{ij} = 1$ if an edge exists between $v_i$ and $v_j$ otherwise $\A_{ij} = 0$. The centrality score $\x_i$ is proportional to the sum of the scores of the vertices connected to it, i.e. 

\begin{eqnarray*} 
\x_i &=& \frac{1}{\lambda} \sum_{v_j \in n(v_i)} \x_j  = \frac{1}{\lambda} \sum_{j=1}^{|V|} \A_{ij} \x_j. 
\end{eqnarray*}

It follows that $\A\x = \lambda \x$ where $\x$ and $\lambda$ are an eigenvector-eigenvalues pair of $\A$, and since all scores must be positive the eigenvector corresponding to the largest eigenvalue of $\A$ gives the vector of scores. 
\end{definition}

One way of finding the largest eigenvector is by using Power Iteration \cite{poole02linalg}. 

\section{Results and Analysis}\label{sec:resultsAnalysis}  

\begin{figure}[h]
\subfigure[Vertices and edges]{\includegraphics[scale=0.3]{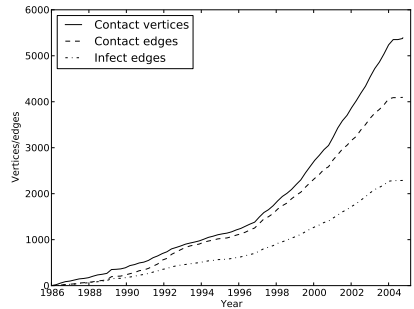}\label{subfig:VE}}
\subfigure[$\log(|V|)$ versus $\log(|E|)$/$\log(|D|)$]{\includegraphics[scale=0.3]{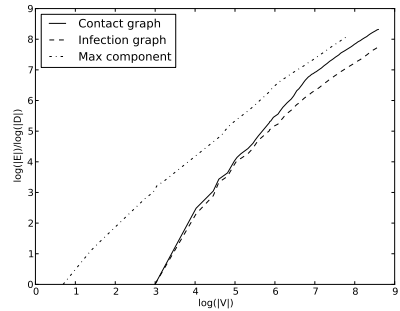}\label{subfig:logVE}} 
\caption{The growth rate of edges and vertices in the graphs.} 
\label{fig:vertexGrowth}
\end{figure}

We start by considering the growth in the number of vertices and edges in the contact and infection graphs, Figure \ref{subfig:VE}. The number of detections per unit time increases over time and there are fewer edges in each graph than vertices implying a low average connectivity.  Since we are using data up until the end of 2004, the end of the plots are flat as not all detected individuals towards the end of this period have been entered into the database. The detection rate increases from 125/year to 430/year after 1997 and there is an increasing discrepancy between the number of vertices and edges in the contact graph. Notice also the sudden increase in detections in 1989. The proportion of contact edges in which an infection occurs decreases: it is 0.426 at 31/12/89 and 0.35 at 31/12/93 indicating more effective contact tracing at the start of the epidemic.

Figure \ref{subfig:logVE} provides more insight into the relationship between the number of vertices and edges in the graphs. A linear regression shows that for the entire contact graph:

\begin{equation*}
\log(|V|)=1.294 \log(|E|)-2.399, 
\end{equation*}

\noindent
$R^2=97\%$, where both regression coefficients are significant. The slope $\sigma$ of the curve of Figure \ref{subfig:logVE} is significant and estimated at $\widehat{\sigma}=1.294$ with the confidence interval $[1.241,1.346]$. This value is closer to 1 than to 2 and later we shall see how this
affects the mean degree. On closer examination the growth exponent is 2.4 at the start of the graph growth and 0.89 at the end, hence rate of growth of the number of contact edges with the number of vertices slows over time and this corresponds to the stabilization of the degree distribution. For the maximum component we observed that the growth exponent was 1.59 at the start and 0.88 at the end of the recorded period. Furthermore the corresponding plot for the edges in the infection graph is close to that of the contact graph at the start of the recorded period and then diverges as expected.

\subsection{Component Distribution} 

\begin{figure}[h]
\subfigure[Number of components]{\includegraphics[scale=0.3]{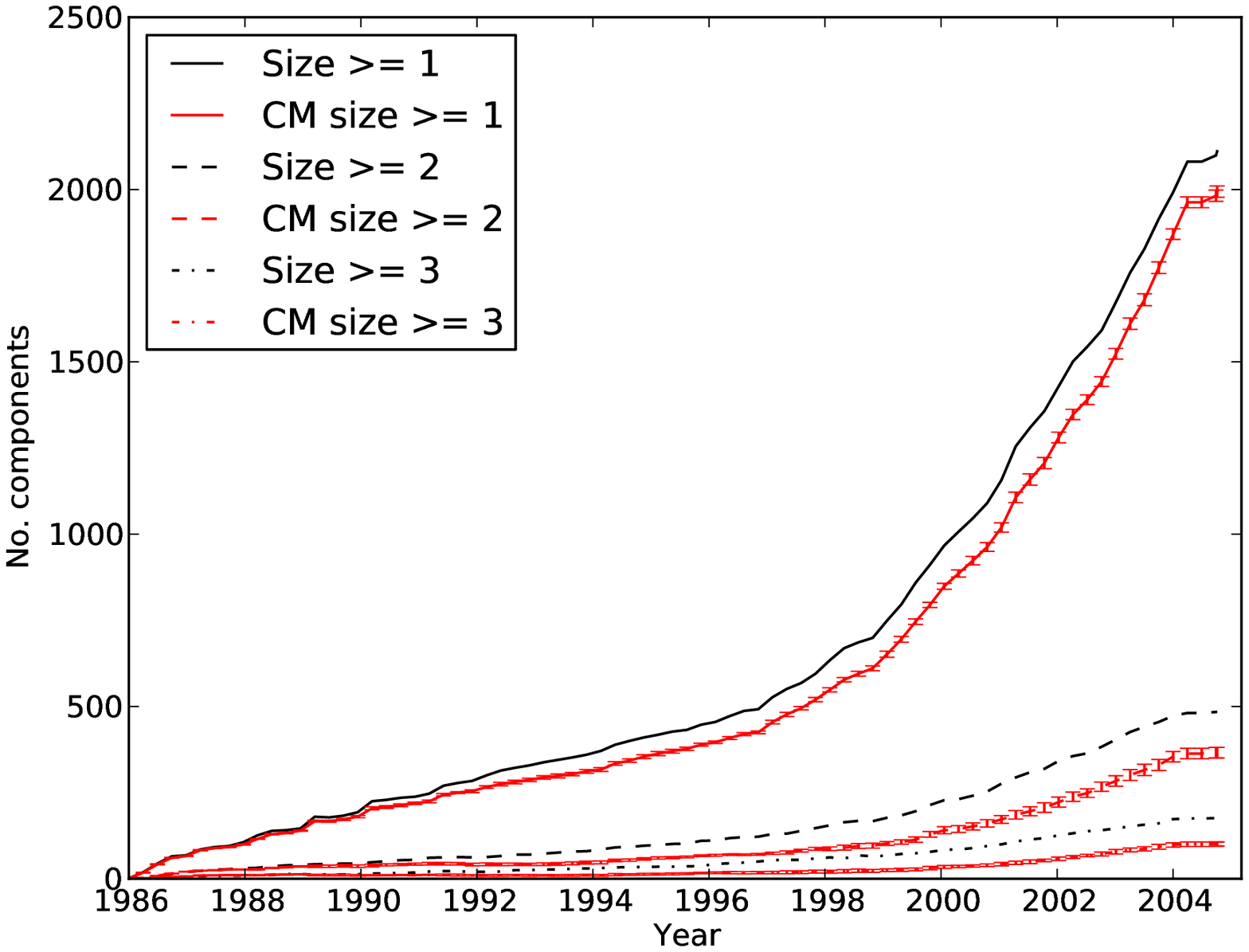} \label{subfig:numConnectedGrowth}} 
\subfigure[Number of trees]{\includegraphics[scale=0.3]{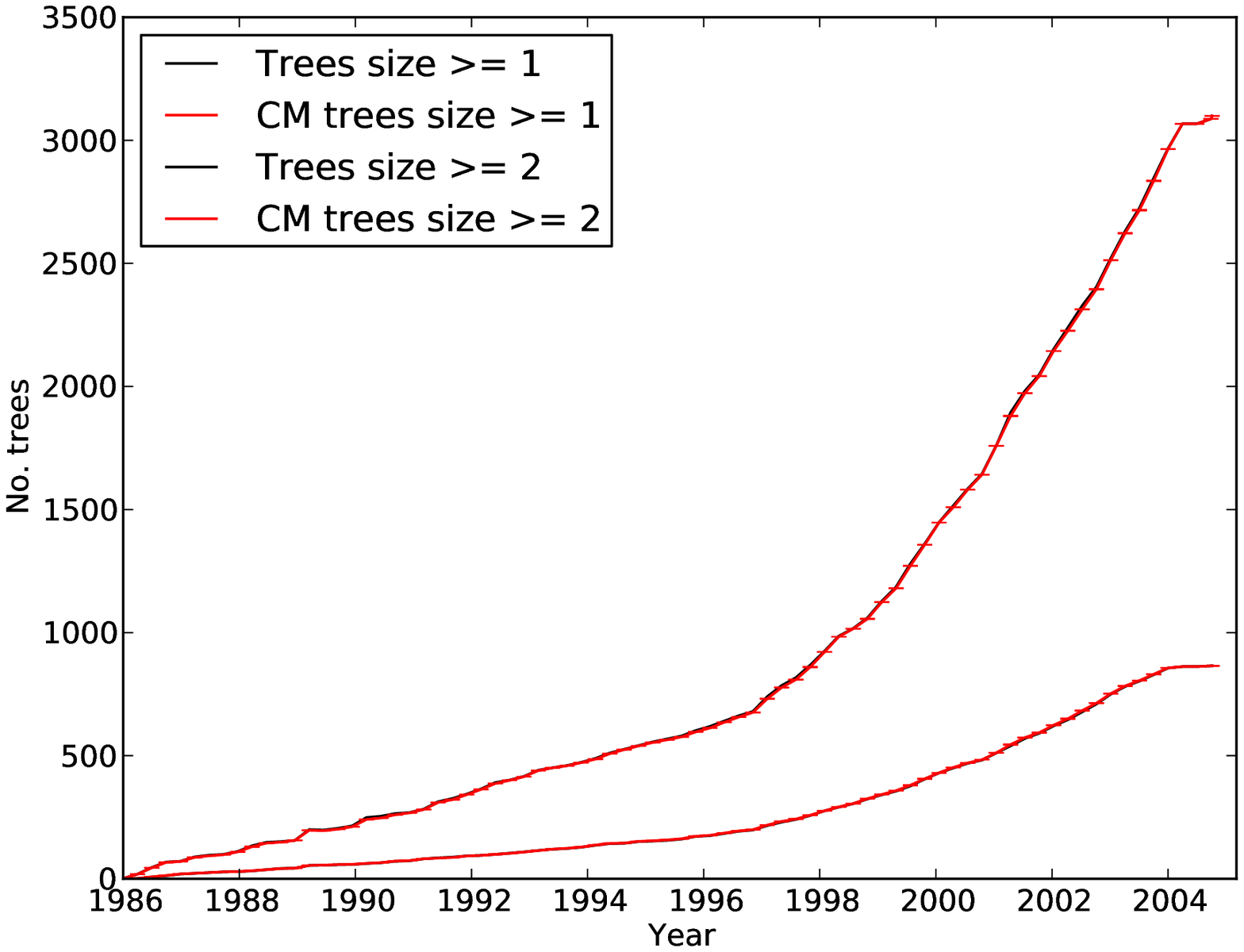} \label{subfig:numTreesGrowth}} 
\caption{The change in the number of components and trees. The difference between the curves for components/trees greater than or equal to 1 and 2 is the number of isolated individuals.  The error bars indicate a difference of 1 standard deviation in the values of interest over the configuration model graphs. } 
\label{fig:componentsDetections}
\end{figure}

One of the reasons for the apparent slow growth of contact edges with respect to vertices is that there are a large number of isolated individuals detected randomly, see Figure \ref{fig:componentsDetections}. There are sudden increases in gradient of the total number of components after 1989, 1990, and to a larger extent after 1997. The increases in 1989 and 1990 correspond to detections of isolated vertices. Before 1997 a new component is formed approximately every 8 days, and afterwards this period becomes 1.5 days. We can also see that the number of components that are bigger than a pair is generally less than half the number of pairs. On examination of the probability distribution of component sizes the probability of an isolated vertex is surprisingly constant around $0.77$ at each 4 year interval. Together, isolated and paired vertices account for approximately 90\% of the total number of components in the graph as it changes over time. In a way this is not surprising since many detections are not made using contact tracing, and furthermore many pairs are husband and wife in which the husband/wife does not disclose additional information about his/her contacts.

Increasingly over time the number of components found using the configuration graphs is less than the equivalent number on the real epidemic. At the end of 2004 there are $2111$ real components compared to $1994$ predicted using the configuration model. Similar patterns are observed for components with sizes $\geq 2$ and $\geq 3$. This implies than a new detection generally forms an edge with a smaller component than that expected by random choice. 

On analysis of the trees present within the infection graph, Figure \ref{subfig:numTreesGrowth}, there is a much higher number of trees than components: $3099$ compared to $2111$ and this is to be expected as not all sexual contact from an infected individual results in transmission of HIV. Ignoring scaling however, the number of trees matches closely the corresponding number of components in the contact graph. Whilst most trees are small we observed that the largest tree sizes were respectively 27, 75, 91, 108 and 120 for the dates 31/12/89, 31/12/93, 31/12/97, 31/12/01 and 02/10/04. Over time approximately 87\% of the trees are pairs or isolated vertices and 96.7\% have a depth of 2 or less. We observed an increase in the number of deep trees between 31/12/97 and 31/12/01, in which 80 and 161 had depths of 2 or larger respectively. At the end of 2004 there were 233 trees of this depth. 

A perhaps surprising artefact of this figure is that the CM predicts closely the number of trees of different sizes. However, some insight into this scenario can be gained by noting that the number of predicted trees cannot be fewer than the number of roots in the set of real trees. Furthermore, the number of isolated vertices will be at least that of the real case, and trees of size 2 or more are highly likely to remain greater than size 2 when generating the CMs.

\begin{figure}[h]
\begin{center} 
\subfigure[Detections by type]{\includegraphics[scale=0.3]{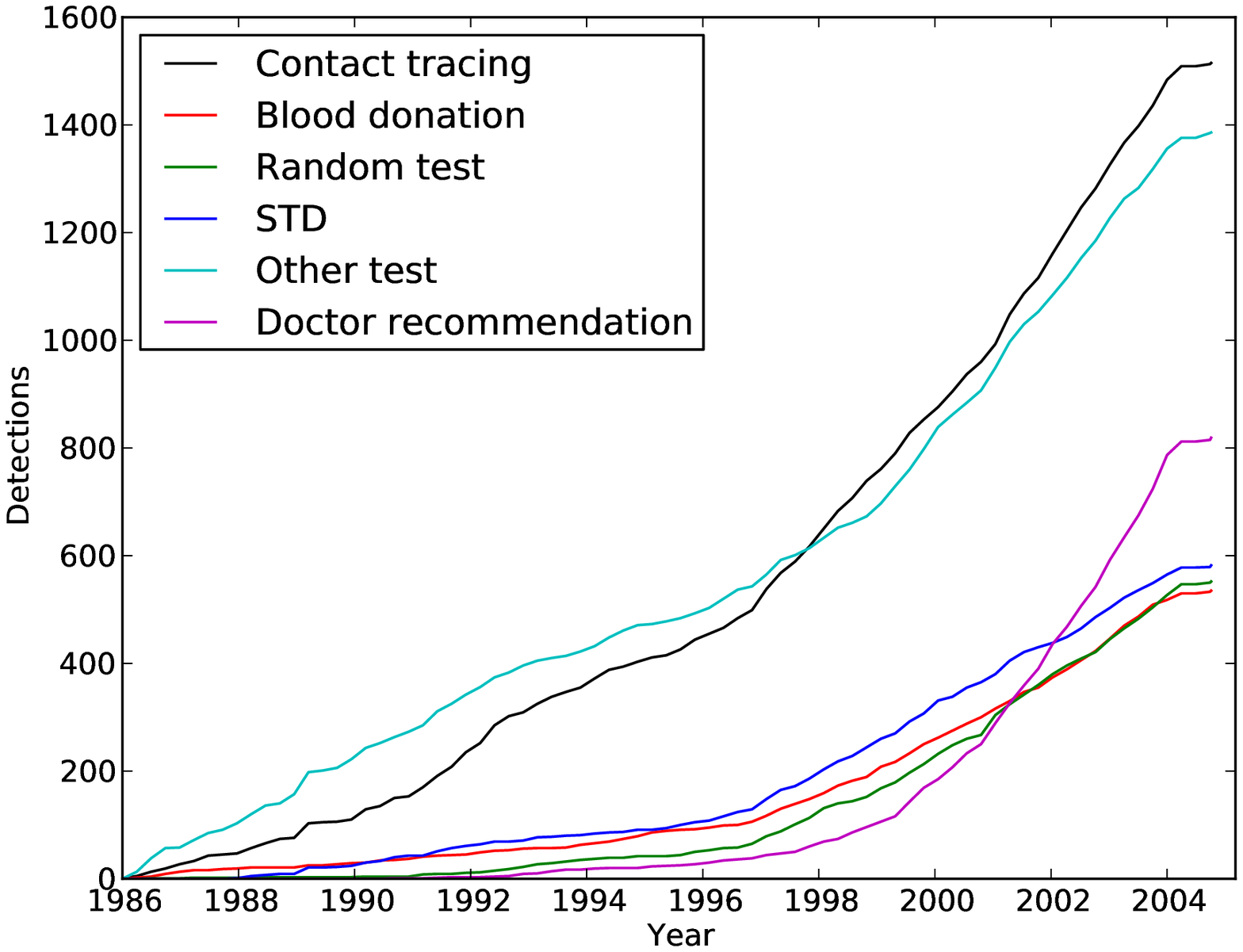} \label{subfig:detectionTypes}} 
\subfigure[Average contacts]{\includegraphics[scale=0.3]{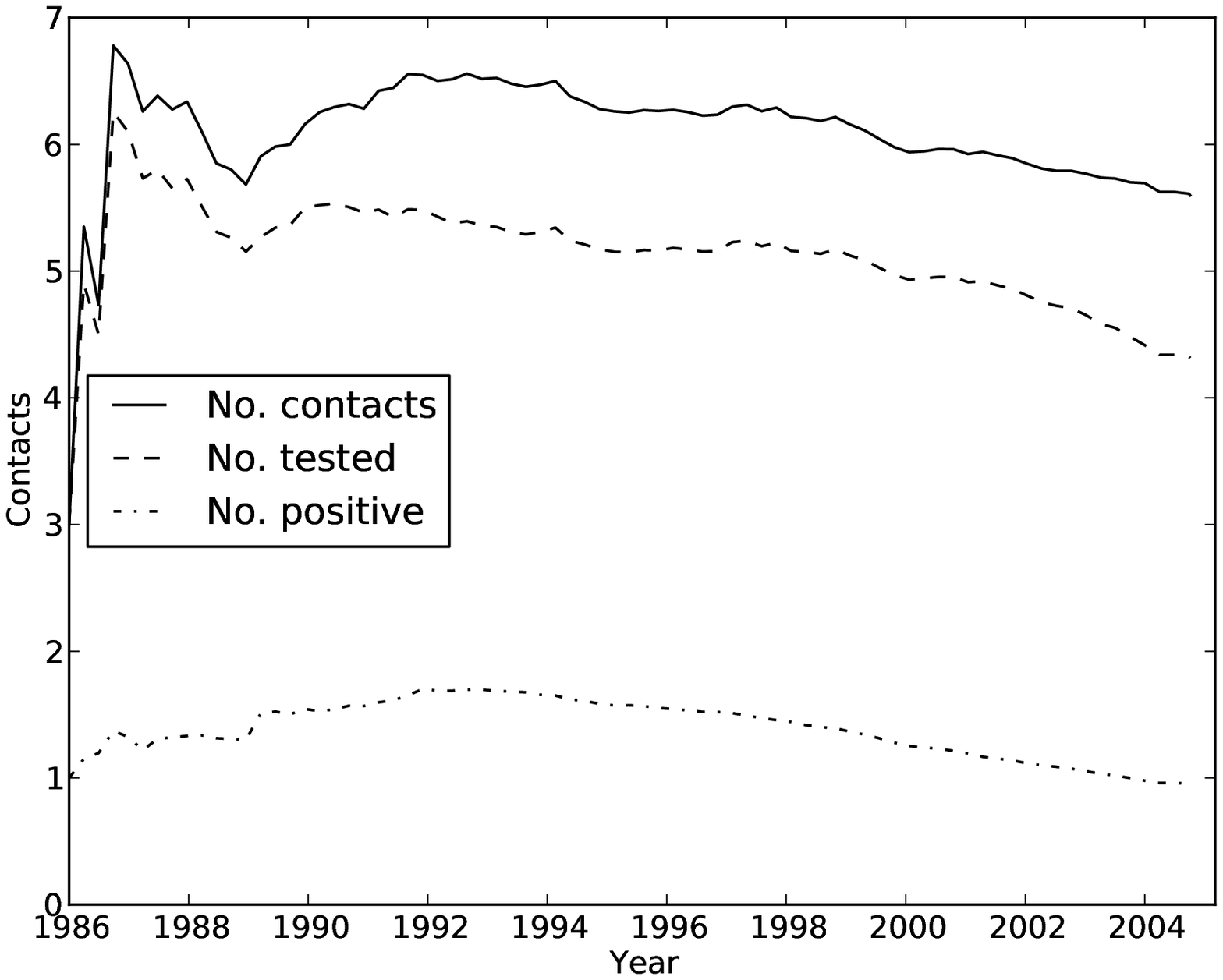}\label{subfig:totalContactsGrowth}}  
\end{center} 
\caption{The total number of detections per year and the change in the mean number of contacts for the two years preceding detection.} 
\label{fig:detections}
\end{figure}

To explore further the trends in the number of components and trees, Figure \ref{subfig:detectionTypes} shows the trends of the most common detection methods. The sudden increase in detections in 1989 can be attributed to increases contact tracing, STD tests and ``other'' tests. There is steady growth in the number of non-contact traced detectees, with doctor recommendations increasing most rapidly after approximately 1997. Contact tracing plays a decreasing role in the overall detection of infected individuals. In general the initial detected epidemic was driven by contact tracing with doctor recommendations and voluntary visits taking an increasing role several years after recording began. The number of detections by voluntary visits is smaller than the detections by doctor recommendations after 2001. The increasing number of non-contact traced detections explains why the number of components in the graph increases exponentially and also the slow rate of edge growth with vertices.

\begin{figure}[h]
\begin{center} 
\subfigure{\includegraphics[scale=0.3]{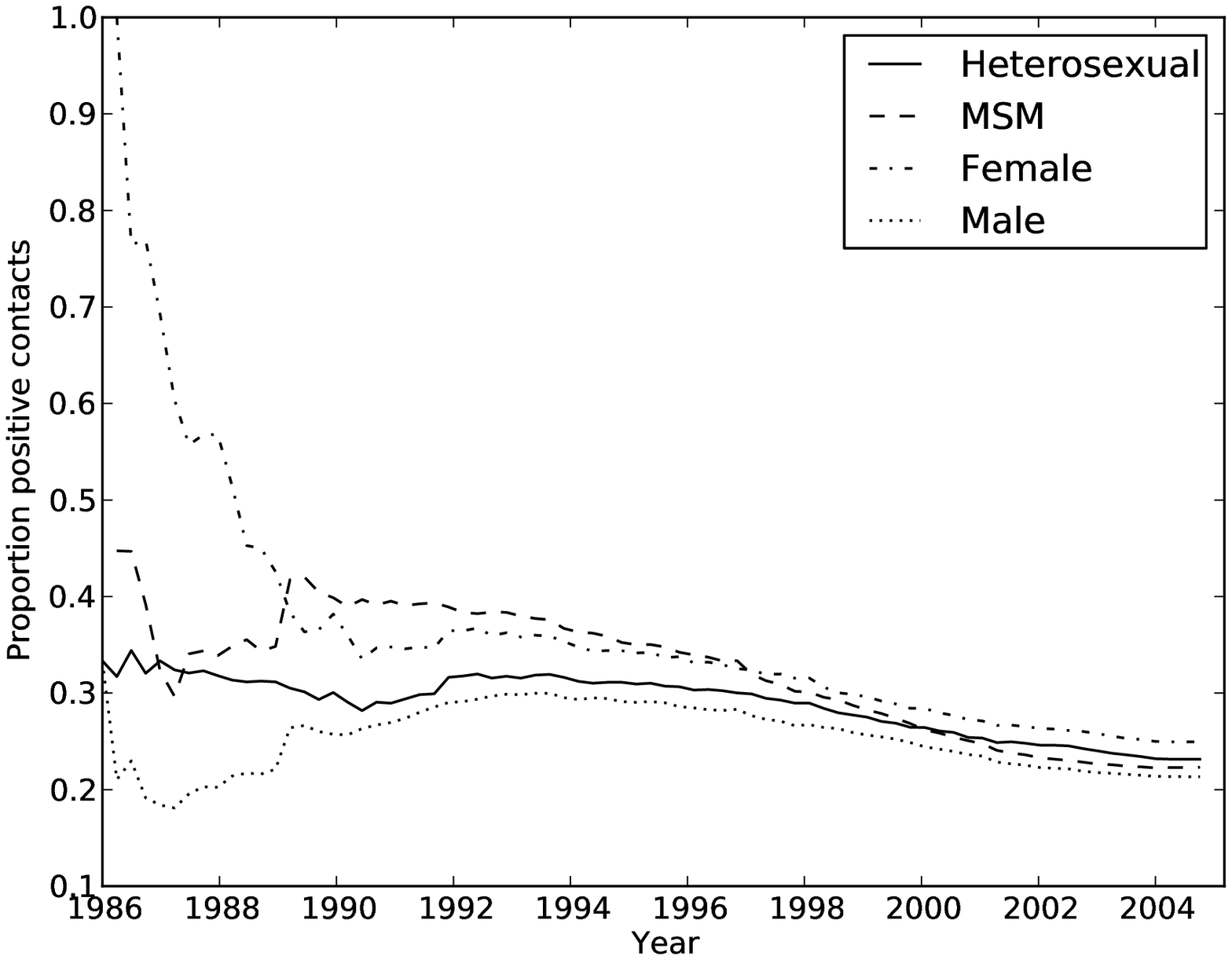}} 
\end{center} 
\caption{The mean proportion of HIV positive individuals of those tested.} 
\label{fig:proportionPositive}
\end{figure}

Also collected in the HIV dataset is the total number of sexual contacts for each individual during the 2 years preceding their detection, and the number of tested and positive individuals in this group. Figure \ref{subfig:totalContactsGrowth} shows the evolution of the mean values of these quantities. The curves for the number of contacts and tested contacts follow similar trends, and encouragingly at least 77\% of contacts are tested on average over time. Both curves have a peak at approximately 1992, at 6.5 and 5.5 respectively, and then decrease which is also reflected in the decrease in the number of positive contacts. At the end of 2004 the mean number of positive contacts is 0.96 per new detectee. The increase in detections in 1989 manifests itself as a lower mean number of contacts and tested contacts although a greater number of positive contacts. This implies that newly detection individuals in this period declared fewer contacts however the ones they did declare resulted in more positive detections than average before this point. A breakdown of the number of contacts by gender and orientation shows that early on in the epidemic it is the MSMs who have the highest mean number of contacts with 22.4 in late 1986 however this number falls to 5.8 in late 2004. The mean number of contacts for females starts at 1 and then rises to 6.2 in 1992 after which it stays constant. Figure \ref{fig:proportionPositive} shows the mean proportion of positive contacts found against the number tested.  It shows that at the start of the recorded epidemic females had their contacts tested positive most often. There is a rise in the mean proportion in 1989 for MSM and men, and another small increase just before the end of 1991. 

\subsection{Mean Degrees}

\begin{figure}[h]
\begin{center} 
\includegraphics[scale=0.3]{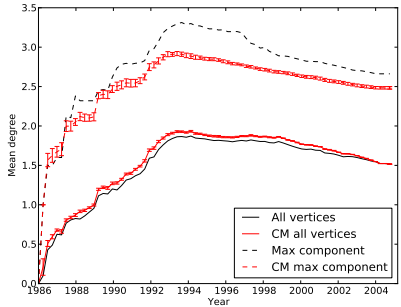}
\end{center} 
\caption{The temporal variation in the average degree.} 
\label{fig:meanDegrees}
\end{figure}

To compare the epidemic evolution to the graphs studied in \cite{leskovec2007graph} we show the mean degrees in Figure \ref{fig:meanDegrees}. The mean out-degrees do not continually rise (as in the graphs in \cite{leskovec2007graph}) and are generally small for both the maximum component and the entire graph. One reason for the peak in the plots is that the non-contact tracing detections starts to become a significant fraction of the total number of detections. The peak mean degrees are 1.8 in 1994 for all vertices and 3.3 for the largest component, after which they decrease slowly. A Chow test shows that there is a breakpoint between 1996 and 1998 for each of these curves. In the first period, there is a regular increase of the mean degree, which can be due to the discovery of individuals with high degrees. After 1994, the mean degree decreases which can be linked to exploration of leaves of the graph, corresponding to individuals with lower degrees. Moreover, an increase in the detection of isolated individuals or couples decreases the mean degree.

For the CM the mean degree for all vertices is slightly higher than that of the observed graphs, $G_1, \ldots, G_T$, at most time points. The reason for this is subtle: in the generation of graphs it is guaranteed that the degree sequence $\d^t$ at time $t$ of $G_t$ is greater than or equal to that of a generated CM graph $\hat{\d}^t$, i.e. $\d^t_i \geq \hat{\d}^t_i$  for all $i, t$. Indeed, we do not explicitly store detection dates in the generated CM graphs, hence using the configuration graph at the final time $T$ to compute the mean degree at time $t$ we simply select the subgraph corresponding exactly to those edges in $G_t$. It is possible that this subgraph has more edges than the equivalent graph $G_t$ and we see it to be the case in Figure \ref{fig:meanDegrees}.

\subsection{Detection Distance Analysis}\label{sectionDetectionDistance}

\begin{figure}[h]
\begin{center}
\subfigure[Giant component]{\includegraphics[scale=0.3]{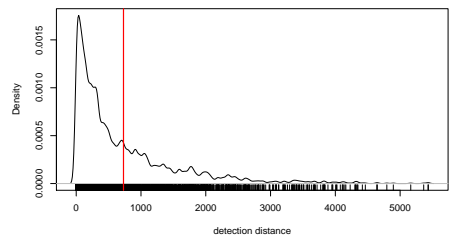}\label{subfig:detectDistGiant}}
\subfigure[All vertices]{\includegraphics[scale=0.3]{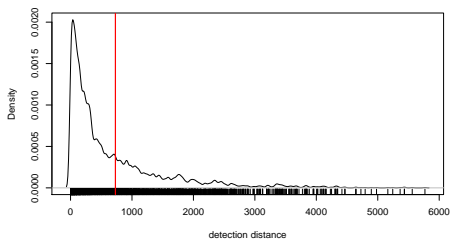} \label{subfig:detectDistAll}}
\end{center}
\caption{Detection distance distributions.} 
\label{fig:detectDists}
\end{figure}

One would hope that the degree of each vertex increases only for a period whilst their contacts are being tested and then stops. Figure \ref{subfig:detectDistGiant} gives the distribution of the detection distance in the largest connected component of the contact graph. 

The distribution can be considered surprising as 1060 edges among 3168 have a detection distance higher than two years, while the official temporal span of the infection tracing procedure is two years. This can be explained partially by the long time the tracing procedure can take in practice, and the continuation of infection after detection.  As shown in Figure \ref{subfig:detectDistAll}, the distribution of the detection distance in the full graph is not very different from what happens in the largest connected component, yet the percentage of high detection distances (more than two years) slightly drops from 0.33 to 0.3.  An interesting question is triggered by this finding: can we relate high detection distance to infection events that occurred after the detection of the infecting person? It turns out that 345 edges in the contact graph connect two persons $a$ and $b$ such that $a$ has been detected more than two years before the probable infection date of $b$. Among those edges, 194 are considered infectious among a total of 2287 infectious edges. 

\subsection{Evolution of the Largest Component} 

\begin{figure}[h]
\begin{center}
\subfigure[Size]{\includegraphics[scale=0.3]{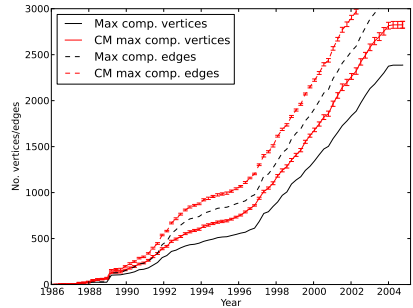} \label{subfig:maxCompSizeGrowth}} 
\subfigure[Mean geodesic distance]{\includegraphics[scale=0.3]{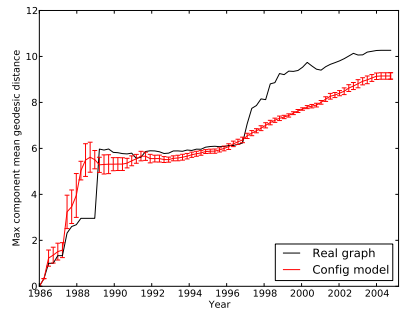} \label{subfig:maxCompGeoGrowth}} 
\end{center}
\caption{The size of the largest component and mean geodesic distances.} 
\label{fig:maxCompGeodesicGrowth}
\end{figure}

We have already touched upon some of the properties of the largest connected component and here we study more of its characteristics. Figure \ref{subfig:maxCompSizeGrowth} shows the number of vertices and edges in the largest component over time. As an aside we note that the size of the 2nd largest component is generally small compared to the largest component being just 18 at the end of the recorded epidemic. Before 1996, the number of nodes of the largest component increases as

$$|V_t|=19.565+8.243 (t-\verb"01/01/1986"),$$
$R^2=99.36\%$, while after 1998,
$$|V_t|=-1784.634+44.125(t-\verb"01/01/1986"), $$

\noindent
$R^2=98.69\%$, which implies a more than 5-fold increase in the rate of new detections within the giant component before 1996 and 1998. It can be verified with a Chow-Fisher test that there is a breakpoint between 1996 and 1998. The curve for the number of edges in the largest component follows almost the same shape as that of the number of vertices. Notice that there is an increase in the number of edges relative to the number of vertices between 1991 and 1992. Before 1991 there are 79 edges per year and afterwards there are 203, an increase by a factor of 2.6. The equivalent CM results show a larger maximum component with more edges although the curves have similar breakpoints. Therefore, in practice a new detection is less likely to connect to the largest connected component than if edges are chosen at random. This corresponds with the trends seen in Figure \ref{fig:componentsDetections}.

\begin{figure}[h]
\subfigure[Hop plot. The initial gradients computed until $k=3$ are $0.313$, $0.460$, $0.447$, $0.423$ and $0.395$ respectively in ascending order of dates.]{\includegraphics[scale=0.3]{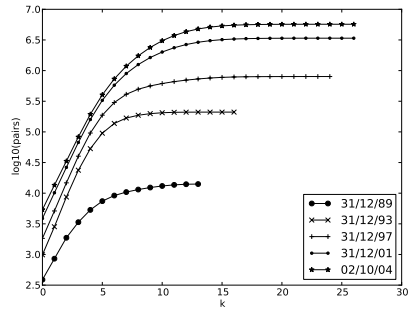} \label{subfig:hopPlot} } 
\subfigure[Effective diameter]{\includegraphics[scale=0.3]{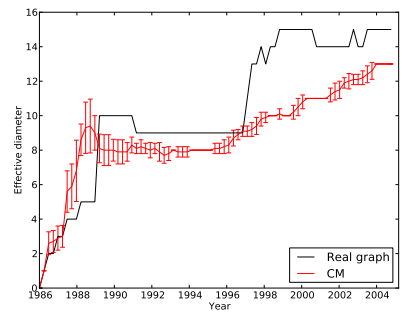} \label{subfig:diameter}} 
\caption{The hop-plot and effective diameter of the largest component over time.} 
\label{fig:diameterGrowth}
\end{figure}

In the next series of analyses we consider several distance measures over the largest component: Figure \ref{subfig:maxCompGeoGrowth} shows the geodesic distance over time, and Figure \ref{fig:diameterGrowth} displays the hop plot and diameter of the largest component. We compared the diameter (not shown) with the effective diameter and found that they both followed broadly similar trends although it is clear that there are long chains in the giant component as the maximum diameter is 33 compared to a maximum effective diameter of 15. The increasing diameter is in direct contrast to the trends observed in the graphs in \cite{leskovec2007graph} and is explained in part due since in general vertices do not continually increase their degrees after detection (see Figure \ref{fig:detectDists}). 

Four periods appear in Figure \ref{subfig:diameter} (and also in Figure \ref{subfig:maxCompGeoGrowth}): two periods of stability in the size of the graph (diameter and geodesic distances) alternate with two periods of expansion of the graph. The diameter rises from 5 to 10 in 1989 and this corresponds to a change in the geodesic distance from 3 to 6. Upon further analysis we can see that in 1989 there is a sudden increase in the largest component size from 25 to 102 and diameter from 5 to 10 in 1989 which implies that the largest component merges with one or more smaller components at this point. This is coupled with a rise in the detection rate during this period. From 1989 to 1997 the diameter and geodesic distances stay constant at 9 (falling from 10 in 1991) and 6 respectively. There were indeed fewer detections during this period due to the economic crisis following the collapse of the Soviet Union. Furthermore, in this period the growth in the largest component is nearly constant at approximately 73 individuals per year. After 1997 this growth rate increases to approximately 256 new detections per year, a significant increase. 

The sudden increases in diameter and geodesic distances at 1989 and 1997 are not reflected in the CM graphs. The diameter and geodesic distances rises earlier from 1987 and remains nearly flat between 1989 and 1995. One intuition is that geodesic distances decrease when there is a ``long distance'' edge between communities and will increase when chains for vertices are added to the largest component. Of note was that the number of vertices of degree 3 or more is 19 in 18/12/88 and rises to 40 in 18/03/89, and this contributes to the peak in the geodesic distance in this period. In the real epidemic the edges formed by this set of vertices do not join communities to the same extent. The number of vertices of degree 1 and 2 continue to increase with the rate of additional degree 1 vertices increasing from 58 to 121 per year after 1997. We already know that the largest component for the CM graphs is generally larger and has more edges than that on the real graphs and this coincidences with the smaller distances after 1997.

The hop-plot in Figure \ref{subfig:hopPlot} shows the extent of degenerate chains of vertices. For example at 31/12/01 the number of pairs reachable with 15 hops is almost identical to those reachable with 27 hops. At $k=0$ the number of reachable pairs is exactly the number of vertices in the graph. The number of reachable pairs of vertices with $k$ hops increases exponentially with $k$ for the initial part of each curve, however the initial gradient at 31/12/89 is lower than the other time points. 

\subsection{Evolution of the Largest Infection Trees} 

In analogy to the study of the largest component, we consider the largest trees in the infection graph. Whereas the largest component was significantly larger than the 2nd largest component, the largest tree is close in size to the 2nd largest tree. Their sizes are respectively 120 and 105 in 2004 which are 5\% and 4.4\% of the size of giant component. Both of these trees continue to grow for nearly the whole recorded period. Observe that the 2nd largest tree before 1999 becomes the largest tree after 1999 and vice versa. Note also the rapid growths in 1989 and 1991 of the largest tree. After 1992 the growth of the largest tree is approximately linear at 6.2 new detectees per year. The CM graphs have much smaller trees of 59.8 and 47.8 at the end of 2004 for the largest and 2nd largest trees however the standard deviations are large.

The tree depths indicate that the 2nd largest tree spawns long chains of vertices, having a depth of 13 at its largest point. This is only slightly shorter than the effective diameter of the largest component. The depth of these trees is larger than that expected for the CM trees after 1989 for the 2nd largest tree and after 1997 for the largest tree. Clearly the large diameters seen in the contact graph correspond to large chains of infections in the infection trees.

As expected, early growth in the largest two trees result from MSM individuals who account for 8 and 10 detections for the 1st and 2nd trees compared to just 2 for heterosexual infectees in 1989. In the period between 1990 and 1991 there is a sudden growth in the number of heterosexual detections for both trees. The MSM population of the 1st tree is comparatively larger than the second at the end of the displayed period. In the following section we show how the trends in sexual orientation manifest themselves in the entire detected population.

\subsection{Gender, Sexual Orientation and Location} 

\begin{figure}[h]
\subfigure[All detection methods]{\includegraphics[scale=0.3]{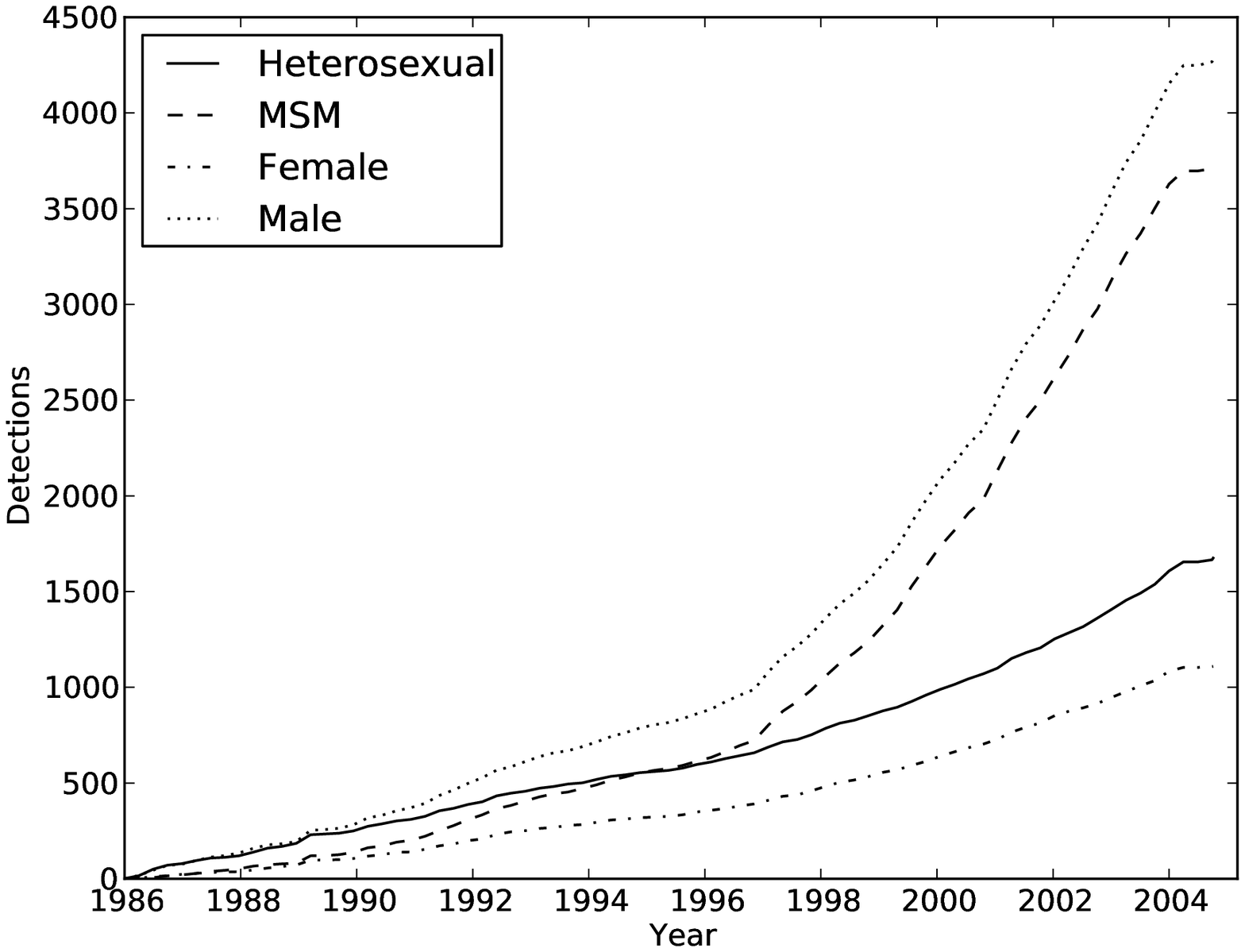} \label{subfig:genderOrientGrowth}}
\subfigure[Contact tracing]{\includegraphics[scale=0.3]{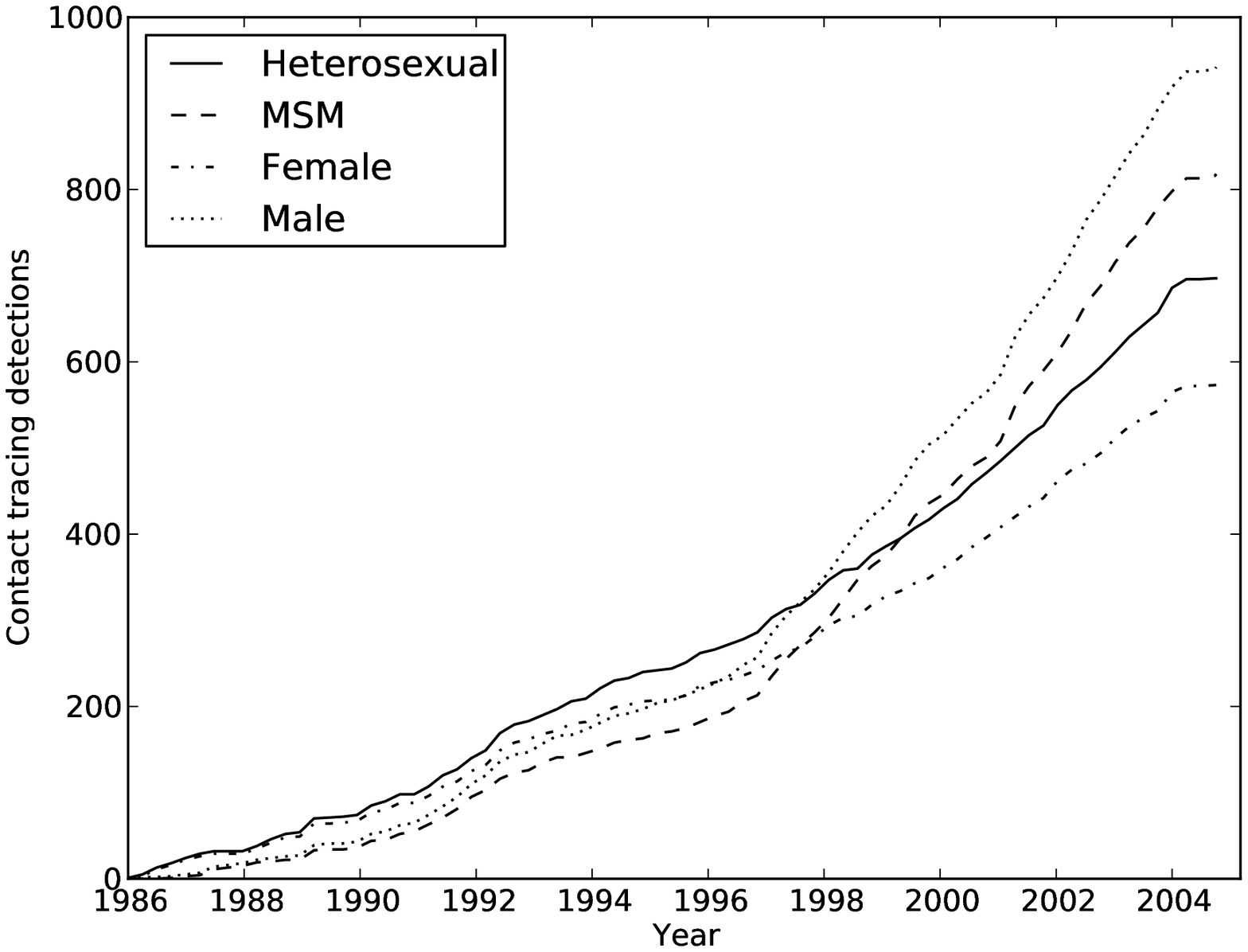} \label{subfig:genderOrientContact}}
\caption{The number of detected individuals by sexual orientation and gender. Again, breakpoints are detected between 1996 and 1998.}
\label{fig:genderOrientLocGrowth}
\end{figure}

We plot the gender and orientation in Figure \ref{subfig:genderOrientGrowth}. After 1996 there are more MSM detections than heterosexual ones and after 1997 the detection rate of MSM individuals (along with males) more than doubles from 108 detections/year to 264 detections/year. In 2002 the MSM detection rate is 497/year in contrast with the heterosexual detection rate of 156/year. These trends explain the rising diameter and geodesic distance starting after 1997. Communities of individuals are newly detected with few ``long distance'' edges which would shorten the diameter and mean geodesic distances. The MSM population starts to grow at a faster rate than the heterosexual one at approximately 1991 with a detection rate of 86/year compared to that of 60/year for heterosexuals. To quantify this increase, the number of newly detected heterosexual males from 13/03/90 to 02/10/04 is 410 whereas the change in the MSM population in the same period is 3552. Women account for a small and decreasing proportion of the contact graph, however a majority of the heterosexual population, 66.2\%, at the end of 2004. A slightly different picture is presented in Figure \ref{subfig:genderOrientContact} on the individuals detected using contact tracing. In this case we see a far smaller difference in the number of people that are MSM versus those that are heterosexual and it is only in 1999 that there are more MSM detections than heterosexual ones. Detection rates before 1997 for all curves are similar. On analysis of the breakdown of individuals detected using doctor recommendations, there are just 5 people detected in late 1992 and it is in 1999 that there is an increase in detection rate for MSM individuals from 23 detections/year to 61 after 1999. By the end of 2004 there are 601 bisexuals detected using this method versus 218 heterosexuals.

\begin{figure}[h]
\begin{center}
\subfigure[Sexual transmission characteristics]{\includegraphics[scale=0.3]{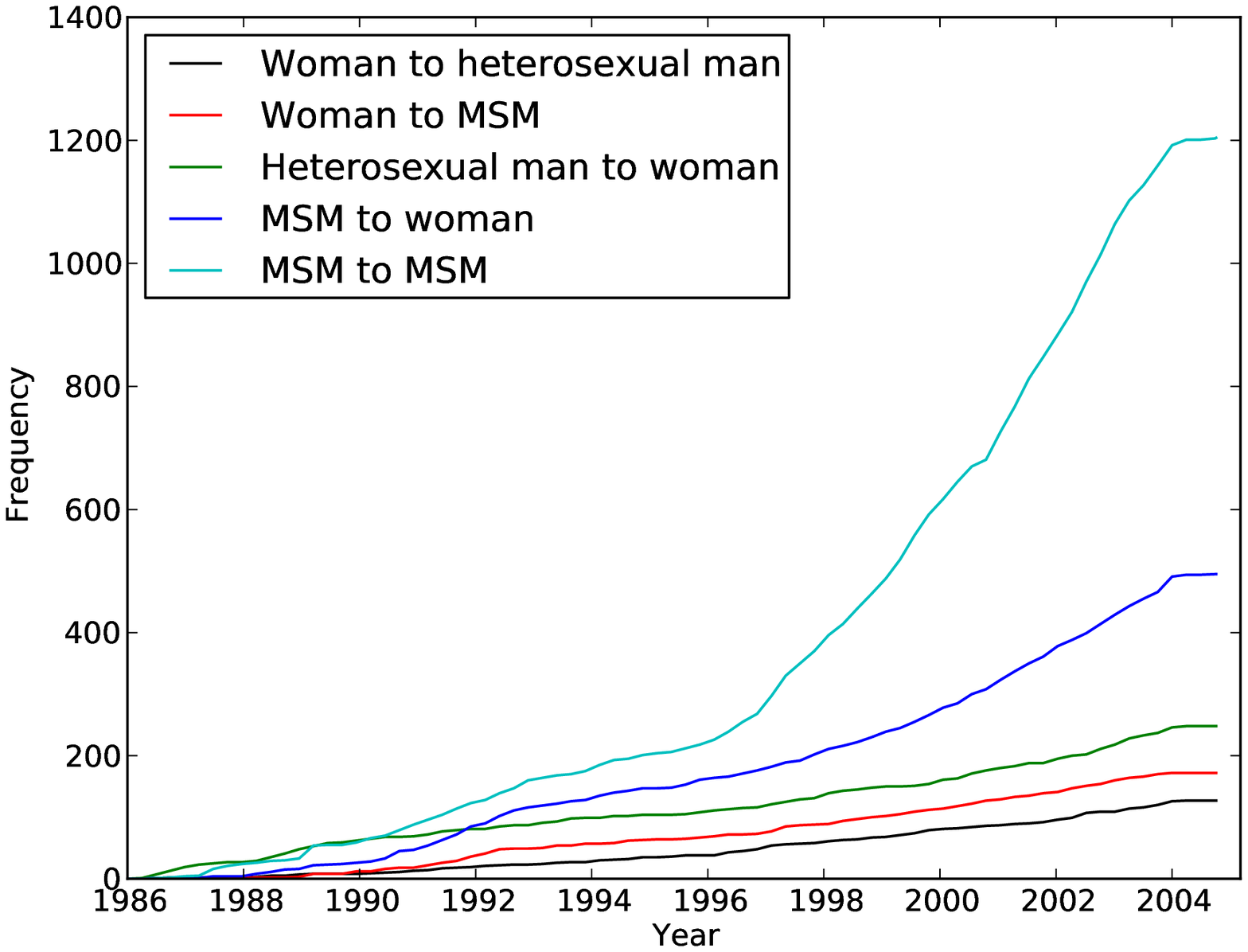}\label{subfig:infectEdgeOrient}} 
\subfigure[Location]{\includegraphics[scale=0.3]{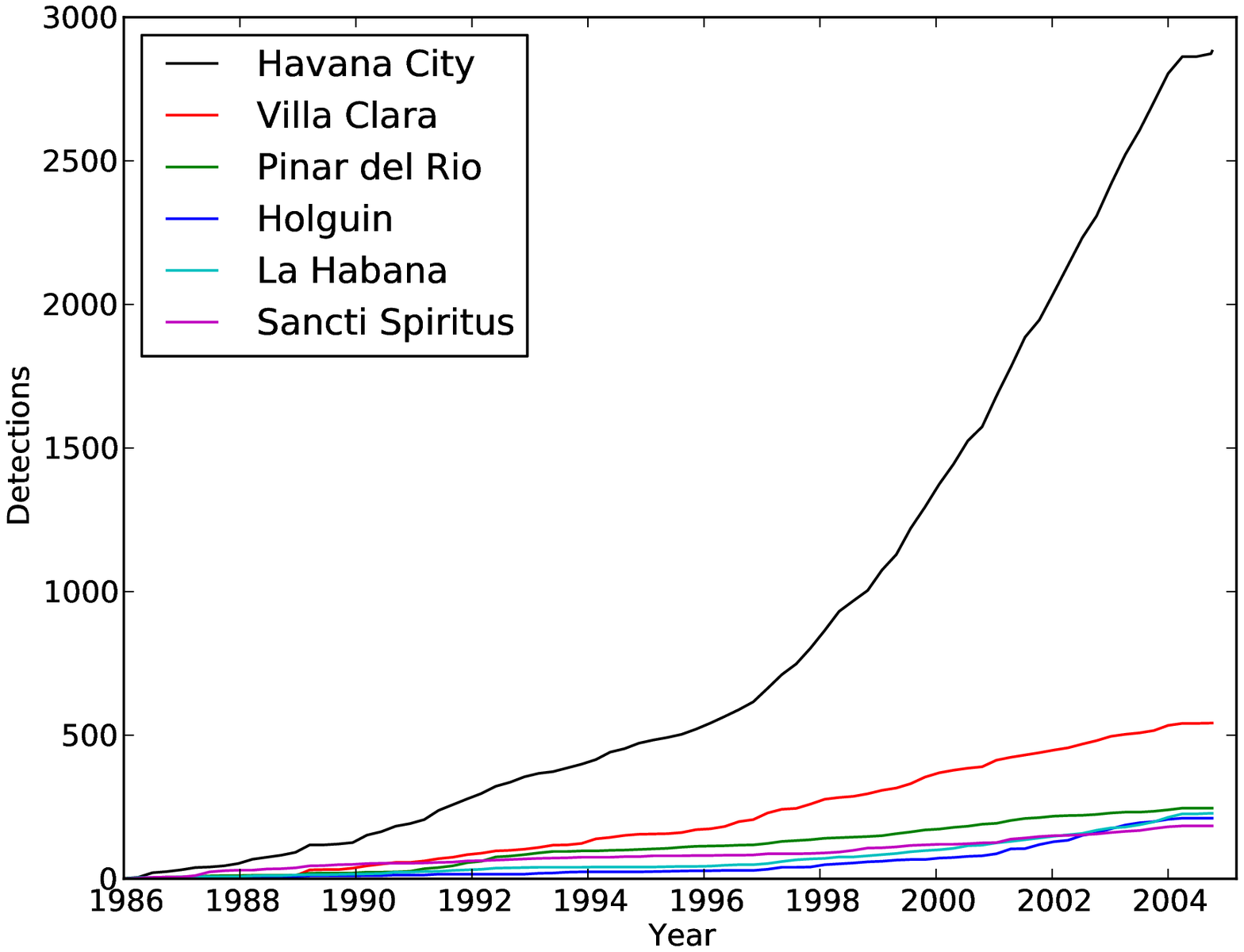} \label{subfig:provinceGrowth}} 
\end{center}
\caption{Gender/orientation of ego and alter along infection edges and the change in the detections by location.} 
\end{figure}

We study next in Figure \ref{subfig:infectEdgeOrient} the frequency of infection based on the gender and sexual orientation of the ego and alter. As one might suspect from previous results the main means of infection occur between MSM individuals and also from MSN to women. The epidemic before 1987 was driven almost exclusively by heterosexual men infecting women. After 1987, there is an increase in MSM to MSM infections and in 1988 we see that infections suddenly rise between MSM individuals and women, and between MSM detectees. In 1990 the infection rates MSM to MSM and MSM to women are 29 and 21 per year. In contrast at 1997 these rates are respectively 102 and 26 infections per year. 

To investigate further the effects of the MSM population in the contact graph, we compute the probability distribution of triangle participation at 4 year intervals. Recall that only MSM contacts can form triangles. Most triangles are seen at 31/12/93 in which 14.3\% of individuals are present in at least 1 triangle, and the number of individuals without any triangles decreases from 31/12/89 to 31/12/93 and then increases. Due to the increase in isolated individuals, the most rapid increase in triangles is between 31/12/89 to 31/12/93 in which the number of individuals with 1, 2 and 3 triangles increases from 12 to 58, 5 to 22 and 3 to 10 respectively, and by the end of 1997 there is 1 individual in 11 triangles. The MSM population is small in 31/12/89 and increases in proportion which would explain this trend. Note that the increase in MSM detection rate from 1997 does not impact the number of individuals in triangles and conversely there are fewer individuals in triangles at this point. One contribution to the trend in the number of individuals with no triangles is the increase in the proportion of random detections. At the end of 2004, 319 individuals participate in at least 1 triangle corresponding to 8.6\% of the MSM population. 

The entropies of the sexual orientations in the infection trees show a strong homogeneity at 31/12/89 with a probability of entropy 0 tree being 0.76 versus an average of 0.65 at the other time points (this compares with the probability of entropy 1 as 0.07 versus 0.27 respectively). This is counter-intuitive given that from Figure \ref{subfig:genderOrientGrowth} the overall mix of sexual orientation is most balanced at this point. It appears that although the epidemic population becomes more MSM later on, it is at an earlier stage that infection trees are more homogeneous. 

Next we recorded the number of detections per province for the 6 provinces with the most detections in Figure \ref{subfig:provinceGrowth}. Unsurprisingly most detections are in Havana City as it is the largest city in Cuba (with approximately 2 million residents out of a total of 11 million). The epidemic starts mostly in Havana with 31 detections at the end of 1986, and between the end of 1986 and 1987 there is an increase in detections from 6 to 30 in Sancti Spiritus. Between the end of 1988 and 1989 the number of detections in Villa Clara increases from 10 to 37, and 1990 and 1992 the detections rate in Pinar del Rio increases from 7 detections/year to 32. The increases in detection rates in these cities correspond to newly discovered communities. Most of the activity however, including the increase in the detections starting from 1997, occur in Havana and this change in detection rate is not seen in other cities. The epidemic in Havana grows at a fast rate because it is difficult to manage however in other cities contact tracing ensures the epidemic spread is under control. 

\subsection{Subgraph/subset Distances and Degrees} 

\begin{figure}[h]
\subfigure[Complete graph and men subgraph.]{\includegraphics[scale=0.3]{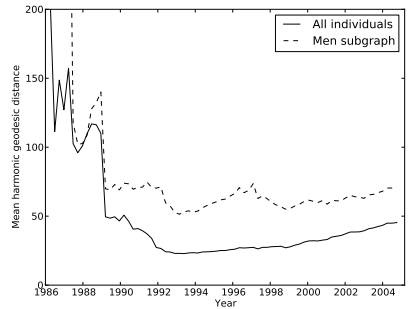} \label{subfig:harmonicAll}} 
\subfigure[Subset of individuals]{\includegraphics[scale=0.3]{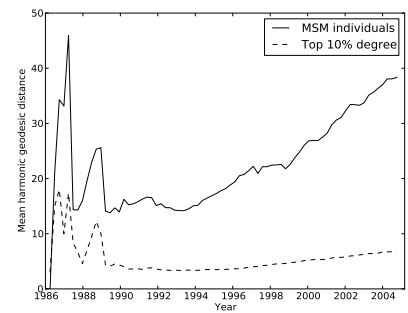}\label{subfig:harmonicTop10}} 
\caption{Harmonic mean geodesic distances.} 
\label{fig:geodesicDists}
\end{figure}

Previously we looked at some distance measures over the contact graph and here we conduct similar analyses for the whole contact graph and subgraphs/subgroups of interest using the harmonic distance. The initial contact graph starts with 1 node and the harmonic distance is undefined. As later individuals are detected using contact tracing, this distance rapidly decreases (see Figure \ref{subfig:harmonicAll}) and fluctuates when components become connected. At 1989 the distance drops from 117 to 50 which corresponds to the merging of two or more components and a sudden increase in detections as previously observed. It is after 1992 that the detection rate in conjunction with contact tracing falls which increases the harmonic distance but does not affect the largest component for example.  

Figure \ref{subfig:harmonicAll} also displays the mean harmonic geodesic distance on the \emph{subgraph} of just men. In this case, connections can only be made between MSM individuals. There is a dramatic fall in distance in 1987 as this corresponds exactly to the point at which MSM to MSM contact starts. There is then an approximately constant mean distance of 71 from 1989 to 1992. The peak at 1997 can be explained by the increase in the detection rate for contact traced MSM individuals at this point. 

In Figure \ref{subfig:harmonicTop10} we consider distances between \emph{subsets} of individuals. For those individuals who are within 10\% of the highest degrees, one observes broadly similar trends to the full detected population except that distances are small in this case indicating a closer connectivity than the general population. The rises and drops in distance before 1989 correspond to those in Figure \ref{subfig:harmonicAll}. The distance rises from under 4 in 1993 to just under 7 in 2004 which is less than the corresponding mean geodesic distances in the giant component of 6 and just above 10. Given the high connectivity of these individuals one would expect that there are many paths between them resulting in low distances. For MSM individuals, distances are lower than the set of all individuals. This said however, during the period 1993 to 2003, the distance changes from 14 to 38 showing that MSM individuals grow further apart at a faster rate than the population as a whole. 

We study the properties of the degree distribution and found them to be broadly constant at 4 year intervals. Note that a degree 1 individual corresponds to someone in a pair or on the periphery of a connected component. We also studied the trend in degree exponents $\alpha$ over time for the set of detected individuals and their complete set of sexual contacts. We noticed that $\alpha$ tends to increase over time from 1.5 to just over 3 in 2004 which implies that the degree distributions are becoming shorter tailed. We also look at the power law exponents of various subsets of individuals according to gender, orientation and detection method. We see that degrees do not vary much across gender or sexual orientation. All the same, after doctor recommendation begins in 1993, it allows one to find individuals who have a slightly higher degree than the ones found by the other methods. Individuals found by random methods have smaller degrees on average.

We also studied the out-degrees of the infection graph. The in-degree of each non-root vertex in this graph is 1 as individuals are infected by a single other person.  As with the contact graph, the curves are similar at 4 year intervals, however at 31/12/89 there is a slightly lower probability of degree 1 vertices. At this point approximately 70\% of vertices have a degree of zero, and 20\% have a degree of one. The degrees therefore have a much more skewed distribution than that of the contact graph. The largest number of infections caused by a single person is 16 compared to a corresponding maximum of 25 detected contacts.  

\subsection{Temporal Analysis of the Clustering}

The clustering obtained in \cite{clemencondearazozarossitran} at the end of the recorded epidemic results in 39 subclusters of the giant-component. The first detected individual of this component was tested seropositive on June 5th 1986. It can be seen that between 1992 and 1995, few clusters are added: the detections in this period mainly contribute to the densification of the existing communities. As a consequence, the diameter does not change much.  The expansion of the graph diameter before 1990 and between 1997 and 1999 correspond to the initial spread of the epidemic and to the discoveries of the new communities which have become infectious between 1989 and 1997 when the research of new infectious individuals was less active.

\begin{figure}[!ht]
\begin{center}
\subfigure{\includegraphics[width=8cm]{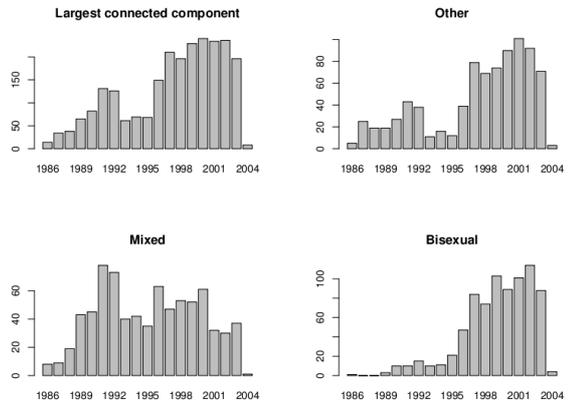} }
\caption{Evolution of the largest component and of the size of its subgroups.} 
\label{fig:evolution-largestandsubgroups}
\end{center}
\end{figure}

A study of their significance in terms of sexual orientation \cite{clemencondearazozarossitran,clemenconarazozarossitranIWANN} shows that 21 out of the latter have atypical distributions: 9 (resp. 12) have less (more) MSM than in the whole population (76\%): they constitute the \textit{mixed} (resp. \textit{MSM}) subgroups. The remaining clusters are called \textit{typical} since their composition looks like the one of the whole population in terms of sexual orientation. The evolution of of the clustering of the giant component is represented in Figure \ref{fig:evolution-largestandsubgroups}. 

We can see that there have been two waves describing the growth of the giant component. This is confirmed by the analysis of the evolution of the subgroups (MSM, mixed and typical) exhibited for the giant component (see Figure \ref{fig:evolution-largestandsubgroups}). We see that between 1986 and 1993, the growth of the giant component is largely due to the subgroups \textit{mixed} and \textit{typical}, then, there is a large increase, after 1995 of all the subgroups sizes, including for the \textit{MSM} subgroup. This latter observation ties in with the trends of Figure \ref{subfig:genderOrientGrowth}. The late contamination of the MSM community can be explained by the fact that the contamination has gone from heterosexual men to women and MSM before propagating among MSM. Because the latter community is relatively well connected, this can explain the higher spread started in the 90's (see \cite{clemencondearazozarossitran}). 

\subsubsection{Temporal Homogeneity} 

\begin{figure}[h]
\begin{center}
\includegraphics[scale=0.3]{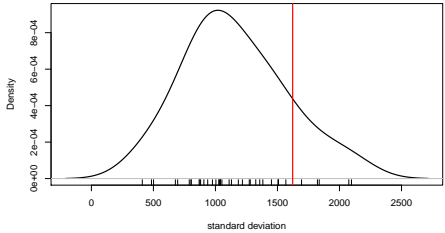}
\end{center}
\caption{Distribution of the standard deviations of the detections dates within the clusters. The red line identifies the global standard deviation of dates.} 
\label{fig:stdCluster} 
\end{figure}

Next we study the temporal homogeneity of the obtained clusters. The detection date of each person is used as the temporal information as this is the only date variable available for each person (as opposed to the infection date, death date, etc.). Detection dates are encoded as a number of days between the earliest detection date in the database and the considered detection date. We study therefore the distribution of this quantity conditioned by the cluster.  The spread of the detection date distributions in each cluster is measured by standard deviations of the dates. Most of the clusters exhibit a large spread, as shown in Figure \ref{fig:stdCluster}. The median standard deviation is 1052.14. That said, most of the clusters are more homogeneous in term of detection date than the full connected component: 34 components out of 39 have a smaller standard deviation than the full component.

\begin{figure}[h]
\begin{center}
\subfigure{\includegraphics[scale=0.3]{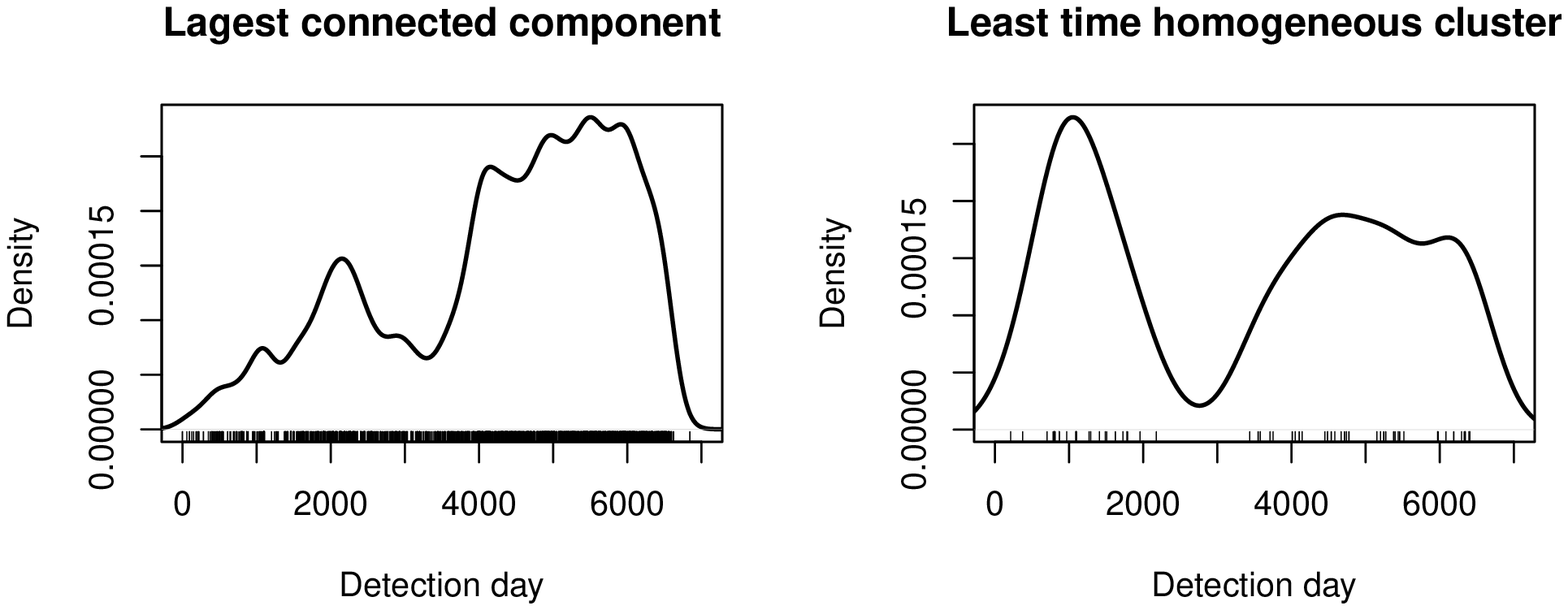}}
\subfigure{\includegraphics[scale=0.3]{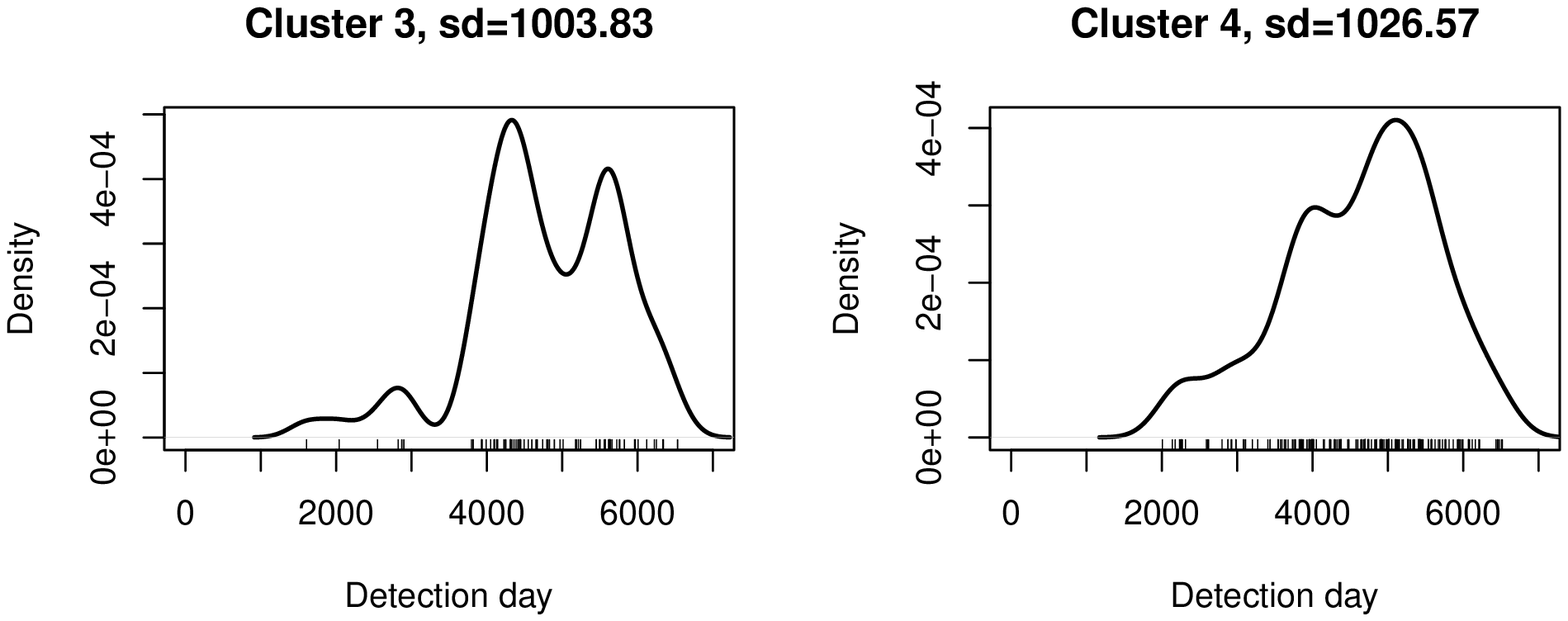}}
\end{center}
\caption{Detection date distributions of the largest component, the least time homogeneous cluster and 2 typical clusters.} 
\label{fig:detectDateCluster}
\end{figure}

Some clusters gather persons detected at extreme dates, as shown in Figure \ref{fig:detectDateCluster}, which compares the distribution of detection days in the full largest connected component to the conditional distribution in the cluster that exhibits the worst time homogeneity. Typical clusters have a standard deviation around 1000 days, roughly three
years. Two examples are presented in Figure \ref{fig:detectDateCluster}. 

Therefore, while the clustering is done without direct knowledge of the temporal information, clusters tend to be homogeneous in term of detection dates compared to the whole network. As indicated by the quite low value of the standard deviation in most of the clusters compared to the global one, this does not happen by chance. We verify this claim with a Monte Carlo approach in which node assignments to clusters are randomized (in a way that preserves the size of the clusters). Then we compute the standard deviations of the detection dates in each cluster and the median of those deviations. The mode of the distribution of this median over 1000 replications is 1615 which is not close to the value obtained by the clustering of 1052.14. 

\subsubsection{Detection Distances}

\begin{figure}[h]
\begin{center}
\includegraphics[scale=0.3]{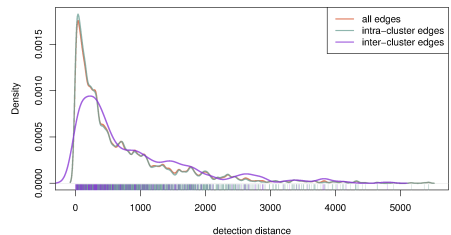}
\end{center}
\caption{Detection distance distributions in conjunction with the clustering.}
\label{fig:detectDistClust} 
\end{figure}

One can also look at temporal homogeneity at the edge level. While it makes sense intuitively to ask to clusters to contain persons detected at the same period, there are numerous examples in the database of connections between persons detected at quite far away dates, as seen in Section \ref{sectionDetectionDistance}. In fact, the detection distance distribution shown in this section partially explains why the clustering exhibits naturally some temporal homogeneity. Indeed, the mode of the distribution is around 40 days and its median is 396. Most of the connections are between close persons (in term of detection distance), and most of the connection between persons occur inside clusters (there is only 333 inter-cluster edges among 3168 edges). Additionally, as shown in Figure \ref{fig:detectDistClust}, inter-cluster edges correspond to higher detection distance. This should reinforce the homogeneity effect.

Differences are confirmed by elementary statistics: the median detection distances between members of distinct clusters is 528 while it is only 379 for members of similar clusters. A simple Monte Carlo study shows a similar trend: only 1 random subset of the edges out of 1000 give a median detection distance as high as the one observed in the intra-cluster edges (random edge subsets have the same size as the intra-cluster edge set).

\subsection{The Most Central Vertices} 

For the final set of analyses we consider those individuals who are most central to the epidemic according to the maximum eigenvector coefficients of the adjacency matrix. Recall from the definition of eigenvector centrality that the coefficient for the $i$th vertex $\x_i = \sum_{j=1}^{|V|} \A_{ij} \x_j$ and hence the sum of the scores of the neighbouring vertices. It follows that in the case a vertex has no neighbours, its score will be zero. The distribution of eigenvector coefficients is long tailed with few vertices having a coefficient larger than 0.15, and the proportion of vertices with this level of centrality decreases rapidly. At 31/12/93 for example 99.19\% of the vertices have a coefficient smaller than 0.15
compared to 99.98\% in 02/10/04. 

\begin{center}
\begin{table*}[ht]
\centering 
\begin{tabular}{ l | l l l l l}
\hline
& 31/12/89 & 31/12/93 & 31/12/97 & 31/12/01 & 02/10/04\\
\hline 
Female & 15.4 & 37.4 & 35.5 & 33.3 & 35.1\\
Male & 84.6 & 62.6 & 64.5 & 66.7 & 64.9\\
\hline
Heterosexual & 17.9 & 39.4 & 39.3 & 39.8 & 41.6\\
Bisexual & 82.1 & 60.6 & 60.7 & 60.2 & 58.4\\
\hline
Contact traced & 41.0 & 60.6 & 62.8 & 57.4 & 56.8\\
Blood donor & 5.1 & 3.0 & 3.3 & 3.4 & 2.8\\
RandomTest & 0.0 & 2.0 & 1.1 & 3.6 & 3.9\\
STD & 20.5 & 9.1 & 9.8 & 9.0 & 8.7\\
Prisoner & 2.6 & 5.1 & 6.6 & 11.6 & 12.6\\
Doctor recommendation & 0.0 & 1.0 & 1.1 & 2.6 & 2.8\\
\hline
Mean age (years) & 22.66 & 20.96 & 21.80 & 22.84 & 23.45\\
\hline
Holguin & 0.0 & 0.0 & 0.0 & 0.3 & 1.5\\
La Habana & 0.0 & 1.0 & 2.7 & 2.6 & 3.7\\
Havana City & 15.4 & 14.1 & 19.1 & 20.7 & 23.2\\
Pinar del Rio & 7.7 & 36.4 & 31.7 & 23.3 & 20.4\\
Sancti Spiritus & 15.4 & 3.0 & 3.8 & 5.4 & 6.9\\
Villa Clara & 48.7 & 37.4 & 35.0 & 37.0 & 34.7\\
\hline
Mean degree & 4.26 & 6.32 & 5.54 & 4.51 & 4.09\\
Std degree & 2.93 & 3.37 & 3.29 & 2.99 & 2.81\\
\hline 
\end{tabular} 
\caption{The characteristics of the 10\% most central vertices. Attributes are represented as percentages in all cases except for age and degree. } 
\label{tab:centralVertices} 
\end{table*} 
\end{center} 

Table \ref{tab:centralVertices} shows the characterisation of 10\% of the vertices with the highest eigenvector coefficients. Between 31/12/89 and 31/12/93 we see that the most central vertices become more female, heterosexual, detected using contact tracing, younger and are more likely to be located in Pinar del Rio and less likely in Sancti Spiritus and Villa Clara. After 31/12/93 the changes are less dramatic and the central population becomes more male, heterosexual, detected less using contact tracing, and older at detection. Notice that before this date, 20.5\% of the central individuals are detected due to other STDs and this figure declines after 31/12/93. These patterns mostly correspond with the findings presented for the entire detected population however the highest proportion of central vertices live in Villa Clara and not Havana City. On analysis of the degree distributions of the central vertices we see as expected that they have higher degrees than the general case.

\section{Conclusions} 

Motivated by an earlier study into the static properties of the Cuban HIV epidemic in \cite{clemencondearazozarossitran}, we examined the evolution of the epidemic from the period beginning in 1986 to the end of 2004 and presented a number of analyses characterising the complex behaviour of the network over time. In response to the questions presented at the start of the study we can say that this epidemic does not follow all of the trends of other networks observed in \cite{leskovec2005graphs, leskovec2007graph}. In particular the growth exponent of the number of edges with vertices is not constant but decreases over time, diameters increase and the mean degrees do not continually increase in both the whole graph and largest connected component. These findings can be explained in large part by the fact that upon detection, additionally detected contacts are typically found within a few years. Furthermore, although contact tracing plays an important role in the detection of infected individuals other methods capture the majority of the detections resulting a strongly disconnected graph with an increasing number of isolated and paired vertices. In contrast to \cite{leskovec2005graphs, leskovec2007graph} the giant component was characterised by large geodesic distances and effective diameters and contains degenerate chains of vertices. Analogously, a study of the infection trees shows that the largest trees grow continuously.

A key comparison we made was between the growth of the epidemic and that of a set of CM graphs generated using the evolving degree sequences of the real epidemic. The CM graphs showed that many properties of the real networks are captured purely in the evolving degree sequences. However, the real epidemic has more components and a smaller giant component than one expected using the CM. Furthermore, after 1997 the mean geodesic distances and effective diameters were smaller in the CM graphs corresponding to the point in which long chains of vertices appear in the largest component. These observations are mirrored in the infection graph, as we saw that the real largest trees were larger and deeper than those found using the CM graphs.

The main points in the temporal behaviour of the epidemic are summarised as follows: in 1986 the spread of the epidemic occurs mainly between heterosexual men and women and it is only later in 1987 that MSM individuals are detected in significant numbers. In 1989 the network changes dramatically: one observes a temporary increase in the detection rate caused mainly by people detected through contact tracing, ``other'' and STD testing. There is a noticeable increase in the size and diameter of the giant component in which smaller components merge into it. This corresponds to a drop in the harmonic mean geodesic distance of the entire graph from 120 to 50. The newly detected people declare fewer contacts than those detected earlier on average, however the MSMs result in a higher proportion of positive contact of those tested at this point. In 1991 many new edges are added to the network between existing vertices due to an increasing contact tracing rate and this causes the diameter of the largest component to fall from 10 to 9. Between 1992 and 1995 existing communities are shown to gain more edges in the clustering of the giant component resulting in a constant diameter. In 1992 we see a peak in the mean number of contacts declared and also those tested for HIV. In the following year the mean harmonic distance between MSM individuals reaches its lowest point, followed in 1994 by the peak mean degree. The top 10\% more central vertices after 1993 are increasingly located in Havana City, however Villa Clara still resides most of the central vertices.  The year 1997 represents another large change in the epidemic: the rate of detection increases from 73 to 256 detections/year and this is driven by an increase in contact traced detections and also doctor recommendations of mostly MSM individuals. This change is reflected in an increase in the size, mean geodesic distance and diameter of the giant component, and also an increase in the proportion of pairs of individuals in which an infection has occurred. One explanation for the breakpoint at 1997 is that the lack of detections at the beginning of the 90's has increased the reservoir of undetected infectious individuals and has shifted the epidemics to another trend after 1997.

\section*{Acknowledgements}

This work was supported by the French Agency for Research under grant ANR Viroscopy (ANR-08-SYSC-016-03) and by AECID project D/030223/10.

\bibliographystyle{plain}
\bibliography{references}

\end{document}